\documentclass[twocolumn]{aastex61}

\usepackage{epstopdf}
\newcommand{\Teff}{\mbox{$T_\mathrm{eff}$}}
\newcommand{\Mjup}{\mbox{$M_\mathrm{Jup}$}}
\newcommand{\Msun}{\mbox{$M_{\odot}$}}

\shorttitle{The Young Substellar Companion ROXs 12~B}
\shortauthors{Bowler et al.}
\begin{document}

\title{The Young Substellar Companion ROX\lowercase{s} 12 B: \\ 
Near-Infrared Spectrum, System Architecture, and Spin-Orbit Misalignment\footnote{Some 
of the data presented herein were obtained at the W.M. Keck Observatory, which is operated as a scientific partnership 
among the California Institute of Technology, the University of California and the National Aeronautics and Space Administration. 
The Observatory was made possible by the generous financial support of the W.M. Keck Foundation.}}

\correspondingauthor{Brendan P. Bowler}
\email{bpbowler@astro.as.utexas.edu}

\author[0000-0003-2649-2288]{Brendan P. Bowler}
\altaffiliation{Hubble Fellow}
\altaffiliation{Visiting Astronomer at the Infrared Telescope Facility, which is operated by the University of Hawaii 
under contract NNH14CK55B with the National Aeronautics and Space Administration}
\affiliation{McDonald Observatory and the Department of Astronomy, The University of Texas at Austin, Austin, TX 78712, USA}

\author{Adam L. Kraus}
\affiliation{McDonald Observatory and the Department of Astronomy, The University of Texas at Austin, Austin, TX 78712, USA}

\author{Marta L. Bryan}
\affiliation{California Institute of Technology, 1200 E. California Blvd., Pasadena, CA 91125, USA}

\author{Heather A. Knutson}
\affiliation{California Institute of Technology, 1200 E. California Blvd., Pasadena, CA 91125, USA}

\author{Matteo Brogi}
\altaffiliation{Hubble Fellow}
\affiliation{Center for Astrophysics and Space Astronomy, University of Colorado at Boulder, Boulder, CO 80309, USA}

\author{Aaron C. Rizzuto}
\affiliation{McDonald Observatory and the Department of Astronomy, The University of Texas at Austin, Austin, TX 78712, USA}

\author{Gregory N. Mace}
\affiliation{McDonald Observatory and the Department of Astronomy, The University of Texas at Austin, Austin, TX 78712, USA}

\author{Andrew Vanderburg}
\altaffiliation{NSF Graduate Research Fellow}
\affiliation{Harvard-Smithsonian Center for Astrophysics, Cambridge, MA 02138, USA}

\author{Michael C. Liu}
\affiliation{Institute for Astronomy, University of Hawai`i at M\={a}noa; 2680 Woodlawn Drive, Honolulu, HI 96822, USA}

\author{Lynne A. Hillenbrand}
\affiliation{California Institute of Technology, 1200 E. California Blvd., Pasadena, CA 91125, USA}

\author{Lucas A. Cieza}
\affiliation{N\'{u}cleo de Astronom\'{i}a, Facultad de Ingenier\'{i}a y Ciencias, Universidad Diego Portales, Av. Ejercito 441, Santiago, Chile}



\begin{abstract}

ROXs 12 (2MASS J16262803--2526477) 
is a young star hosting a directly imaged 
companion near the deuterium-burning limit.
We present a suite of spectroscopic, imaging, and time-series observations to 
characterize the physical and environmental properties of this system.
Moderate-resolution near-infrared spectroscopy of ROXs~12~B from
Gemini-North/NIFS and Keck/OSIRIS 
reveals signatures of low surface gravity including weak alkali absorption lines and a triangular
$H$-band pseudo-continuum shape.
No signs of Pa$\beta$ emission are evident.
As a population, however, we find that 
about half (46~$\pm$~14\%) of young ($\lesssim$15~Myr) companions with masses 
$\lesssim$20~\Mjup \ possess actively accreting subdisks detected via Pa$\beta$ line emission,
which represents a lower limit on the prevalence of circumplanetary disks in general 
as some are expected to be in a quiescent phase of accretion. 
The bolometric luminosity of the companion and age of the host star (6$^{+4}_{-2}$~Myr) imply a mass of 
17.5~$\pm$~1.5 \Mjup \ for ROXs 12 B based on hot-start evolutionary models.  
We identify a wide (5100 AU) tertiary companion to this system, 2MASS J16262774--2527247, 
which is heavily accreting and exhibits stochastic variability in its $K2$ light curve.
By combining $v$sin$i_*$ measurements with rotation periods from $K2$, we constrain the
line-of-sight inclinations of ROXs 12 A and 2MASS J16262774--2527247 and find that
they are misaligned by 60$^{+7}_{-11}$$^{\circ}$.
In addition, the orbital axis of ROXs~12~B is likely misaligned 
from the spin axis of its host star ROXs~12~A,
suggesting that ROXs 12 B 
formed akin to fragmenting binary stars or in an 
equatorial disk that was torqued by the wide stellar tertiary.

\end{abstract}


%\keywords{editorials, notices --- 
%miscellaneous --- catalogs --- surveys}



\section{Introduction}{\label{sec:intro}}

A growing number of directly imaged planetary-mass companions (PMCs; $\lesssim$13~\Mjup) spanning orbital distances of hundreds to thousands of AU 
have been discovered over the past decade through adaptive optics imaging and seeing-limited common proper motion
searches (e.g., \citealt{Ireland:2011id}; 
\citealt{Bailey:2014et}; \citealt{Naud:2014jx}; \citealt{Kraus:2014tl}; \citealt{Deacon:2016dg}; \citealt{Bowler:2017hq}).  
The origin of this remarkable population is under debate but probably differs from 
giant planets located at smaller separations within $\sim$10 AU.
Planet scattering to large separations (\citealt{Boss:2006ge}; \citealt{Gotberg:2016ge}), 
a binary-like formation scenario via turbulent fragmentation 
(\citealt{Low:1976wt}; \citealt{Boss:2001vw}; \citealt{Bate:2009br}),
and instabilities in massive protoplanetary disks have all been proposed to explain their existence.  
Unfortunately, robust tests of these scenarios are difficult with the relatively small sample of known objects 
(\citealt{Bowler:2016jk}).

In principle,
clues about their origin can be inferred from a comparative abundance analysis with their host stars 
(e.g., \citealt{Konopacky:2013jvc}, \citealt{Barman:2015dy}),
measurements of their occurrence rate over time, long-term monitoring of their orbits 
(e.g., \citealt{Ginski:2014ef}; \citealt{Rameau:2016dx}; \citealt{Blunt:2017eta}), 
the physical properties of their circumplanetary disks (\citealt{Kraus:2015fx}; \citealt{Bowler:2015hx}; \citealt{Stamatellos:2015dx}), 
and constraints on their mass and semi-major axis distributions (\citealt{Biller:2013fu}; \citealt{Brandt:2014cw}; \citealt{Reggiani:2016dn}).
Indeed, \citet{Bryan:2016eo} recently concluded that dynamical scattering is probably not the dominant origin of wide PMCs
based on the lack of close-in scatterers, the low rate of close-in giant planets in the field,
and early orbital constraints for wide PMCs.

Regardless of their formation route, the favorable angular separations and contrasts of wide PMCs 
make them attractive targets for detailed spectroscopic characterization in the near-infrared 
(e.g., \citealt{Bowler:2014dk}; \citealt{Gauza:2015fw}).
As such, they represent excellent targets to study the atmospheres of young 
gas giant planets and serve as empirical templates for discoveries with second-generation 
planet-finding instruments like the Gemini Planet Imager, SPHERE, and SCExAO.

\citet{Kraus:2014tl} presented the discovery of  faint companions comoving with 
the pre-main sequence stars FW Tau AB, ROXs 42B, and ROXs 12 at projected separations between 100 to 400~AU.  
At the young ages of these systems (1--10~Myr), the implied masses of the companions
span about 5--20~\Mjup \ assuming hot-start cooling models.
Follow-up spectroscopy by \citet{Bowler:2014dk} showed that ROXs 42B b is a young early L dwarf
with clear spectroscopic signs of low surface gravity.
On the other hand, the companion to FW Tau exhibits a mostly featureless near-infrared pseudo-continuum spectrum with strong veiling 
and emission lines indicating ongoing accretion and outflow activity.  The implication is that the faint companion to FW Tau may be a brown dwarf with 
an edge-on disk rather than a widely-separated planet, as suggested from photometry alone.
Follow-up spectroscopic confirmation is evidently a critical step in 
confirming the low temperatures and masses of young planet candidates found with direct imaging.

Here we present near-infrared integral-field spectroscopy of the substellar companion to ROXs 12
(2MASS J16262803--2526477), an M0-type pre-main sequence star\footnote{There 
has been some confusion regarding the coordinates of ROXs 12 in the literature.
\citet{Montmerle:1983jc} originally identified 47 ``Rho Oph X-ray'' (ROX) detections in observations with the \emph{Einstein Observatory} X-ray telescope.
ROX~12 is listed at $\alpha_\mathrm{J2000.0}$ = 16:26:24.8, $\delta_\mathrm{J2000.0}$ =  --25:27:28 with a 40$''$
positional uncertainty.  
In a follow-up study to identify the sources of the X-ray emission,
 \citet{Bouvier:1992ww} list ROXs 12 (``ROX star 12''), the candidate optical counterpart to the ROX 12 detection,
 as the star at $\alpha_\mathrm{J2000.0}$ = 16:26:28.0, $\delta_\mathrm{J2000.0}$ =  --25:26:47
 (also known as 2MASS J16262803--2526477).  Later, \citet{Ratzka:2005jv} used speckle imaging to identify
 a candidate companion to ROXs 12 at  a position angle of 10.3$^{\circ}$, a separation of 1$\farcs$75, and a flux ratio of $\Delta K$=5.7 mag, but the 
 coordinates they list are for a \emph{different} star at $\alpha_\mathrm{J2000.0}$ = 16:26:27.75, $\delta_\mathrm{J2000.0}$ =  --25:27:24.7
 (also known as 2MASS J16262774--2527247).  \citealt{Kraus:2014tl} obtained followup adaptive optics 
 imaging of ROXs 12 (2MASS J16263803--2526477); while constructing the target list, they
used coordinates based on the 2MASS source closest to the position in
\citet{Bouvier:1992ww}, observing the correct source. However,
in the manuscript that confirmed comovement for the companion
identified by Ratzka et al., they adopted the same (incorrect)
coordinates for the star 37$''$ south of ROXs 12 that were listed by
Ratzka et al. 
Most of the observations we collected between 2011 and 2015 targeted the erroneous published coordinates for ROXs~12,
 but have nevertheless provided useful data for the third companion 2MASS J16262774--2527247.} 
 near the boundary between the Ophiuchus and Upper Scorpius star-forming regions.
ROXs 12~B\footnote{Here we adopt ``B'' rather than ``b'' as used in the discovery paper because the mass we infer for this companion is securely above the deuterium-burning limit.}
 was first identified as a candidate companion by \citet{Ratzka:2005jv} at 1$\farcs$78 (240~AU in projected separation) 
and confirmed to be comoving with its host star by \citet{Kraus:2014tl}.
More recently, \citet{Bryan:2016eo} obtained follow-up high-contrast imaging of this system and found that ROXs 12 B has undergone 
measurable orbital motion based on astrometry spanning about 15 years.
As one of only a handful of young ($<$10 Myr) very low-mass companions at wide orbital distances beyond 100~AU,
ROXs 12~B offers a valuable opportunity to study the atmosphere of a young object spanning the 
brown dwarf-planetary mass boundary.

\begin{figure}
  \vskip -.3 in
  \hskip -1.05 in
  \resizebox{5.4in}{!}{\includegraphics{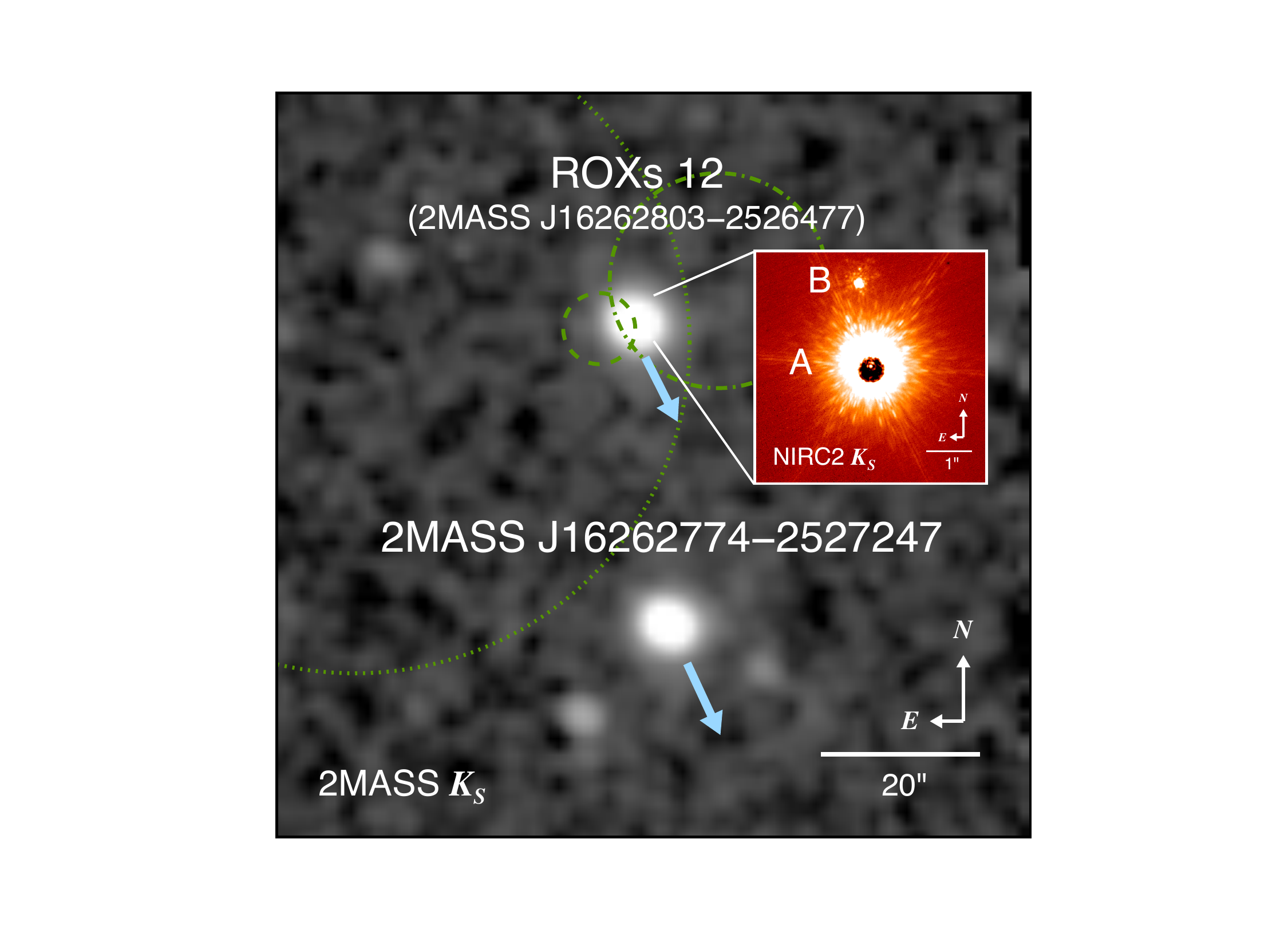}}
  \vskip -.2 in
  \caption{Overview of the ROXs 12 triple system.  This seeing-limited 2MASS $K_S$-band image 
  shows the wide binary ROXs 12 A (2MASS J16262803--2526477)
  and 2MASS J16262774--2527247 separated by 37$''$ (5100 AU).  The substellar companion ROXs 12 B is located 1.8$''$ (240 AU)
  from its host star, seen here in the inset $K_S$-band adaptive optics image 
  from Keck/NIRC2.  
  Green circles show X-ray error circles from \emph{Einstein Observatory} (dotted line; \citealt{Harris:1996uk}),
  $ROSAT$ (\citealt{Voges:2000wn}), and 
  $Swift-XRT$ (dashed; \citealt{Evans:2014cp}).
  Proper motion vectors for ROXs 12 A and 2MASS J16262774--252724 are shown as blue arrows indicating
  the direction of motion (see Section~\ref{sec:binary}); the length of the arrows have been magnified to make them visible.
 North is up and east is to the left. \label{fig:skyimg} } 
\end{figure}

In addition to characterizing the atmosphere of ROXs 12 B, we also show that  
the young disk-bearing star 2MASS J16262774--2527247 (hereinafter 2M1626--2527) located 37$''$ south of ROXs 12 is 
likely a wide binary companion based on a consistent radial velocity and proper motion with ROXs 12.
Figure~\ref{fig:skyimg} shows an overview of the system.
The combination of rotation periods from $K2$ together with projected rotational velocities from our
high-resolution spectra allow us to assess the mutual inclinations of both stars.
Combining these results with orbital inclination constraints for ROXs 12 B allows us to then examine
whether the ROXs 12 system is in spin-orbit alignment, as expected for a planet forming in a disk.
To place ROXs 12 B in context, we conclude with an overview of  
accreting subdisks around young planetary-mass companions and derive
the frequency of accreting circumplanetary disks from 
companions with existing moderate-resolution $J$-band spectra.

% Figure 1

\begin{deluxetable*}{lcccccccc}
\renewcommand\arraystretch{0.9}
\tabletypesize{\small}
\setlength{ \tabcolsep } {.1cm} 
\tablewidth{0pt}
\tablecolumns{9}
\tablecaption{Spectroscopic Observations\label{tab:specobs}}
\tablehead{
        \colhead{Object}   &  \colhead{Date}      &   \colhead{Telescope/}  &  \colhead{Filter}  &  \colhead{Slit Width}  & \colhead{Plate Scale}   &    \colhead{Tot. Exp.}   &   \colhead{Resolution} & \colhead{Standard\tablenotemark{a}}   \\
                     \colhead{}  &  \colhead{(UT)}   &   \colhead{Instrument}       &  \colhead{}           &  \colhead{($''$)}               & \colhead{(mas pix$^{-1}$)}             &    \colhead{(min)}   & \colhead{(=$\lambda$/$\delta \lambda$)}  &   \colhead{}             
        }   
\startdata
ROXs 12 B           &   2016 April 18+22  &  Gemini-North/NIFS  &  $J$   &  $\cdots$  &  100$\times$40   &  110  & 6000  &  HD 145127  \\
ROXs 12 B           &   2016 May 22     &   Keck/OSIRIS  &  $Kbb$   &  $\cdots$  &  50   &  50  & 3800  &  HIP 69021  \\
ROXs 12 B           &   2016 May 22     &   Keck/OSIRIS  &  $Hbb$   &  $\cdots$  &  50   & 60   & 3800  &  HIP 93691  \\
ROXs 12 B           &   2016 May 22     &   Keck/OSIRIS  &  $Jbb$   &  $\cdots$  &  50   &  90  & 3800  &  HIP 93691  \\
ROXs 12 A            &  2011 June 28    &   Keck/HIRES  &  $KV418$   &  0.861  &  $\cdots$  & 5  &  48000  &  $\cdots$  \\
ROXs 12 A            &  2016 July 26    &   McDonald 2.7 m/IGRINS  &  $\cdots$   &  0.98  &  $\cdots$  & 30  &  45000  &  HD 155379  \\
2M1626--2527    &    2011 Apr 29     &   IRTF/SpeX  &  $\cdots$ &   0.3  &  $\cdots$   &   8  & 2000  &  HD 144925  \\
2M1626--2527    &    2014 May 21     &   Mayall/RC-Spec  &  $GG 495$ &   1.5  &  $\cdots$   &   20  & 2600  &  HZ 44  \\
2M1626--2527      &  2015 May 03    &   Keck/HIRES  &  $GG475$   &  0.861  &  $\cdots$  & 10  &  48000  &  $\cdots$  \\
2M1626--2527     &  2016 July 26    &   McDonald 2.7 m/IGRINS  &  $\cdots$   &  0.98  &  $\cdots$  & 30  &  45000  &  HD 155379  \\
\enddata
\tablenotetext{a}{Telluric or radial velocity standard.}
\end{deluxetable*}

\section{Observations}{\label{sec:observations}}

\subsection{Gemini North/Near-Infrared Integral Field Spectrometer $J$-Band Spectroscopy of ROXs 12 B}{\label{sec:nifs}}

\begin{figure}
  \vskip -1 in
  \hskip -2.7 in
  \resizebox{8.2in}{!}{\includegraphics{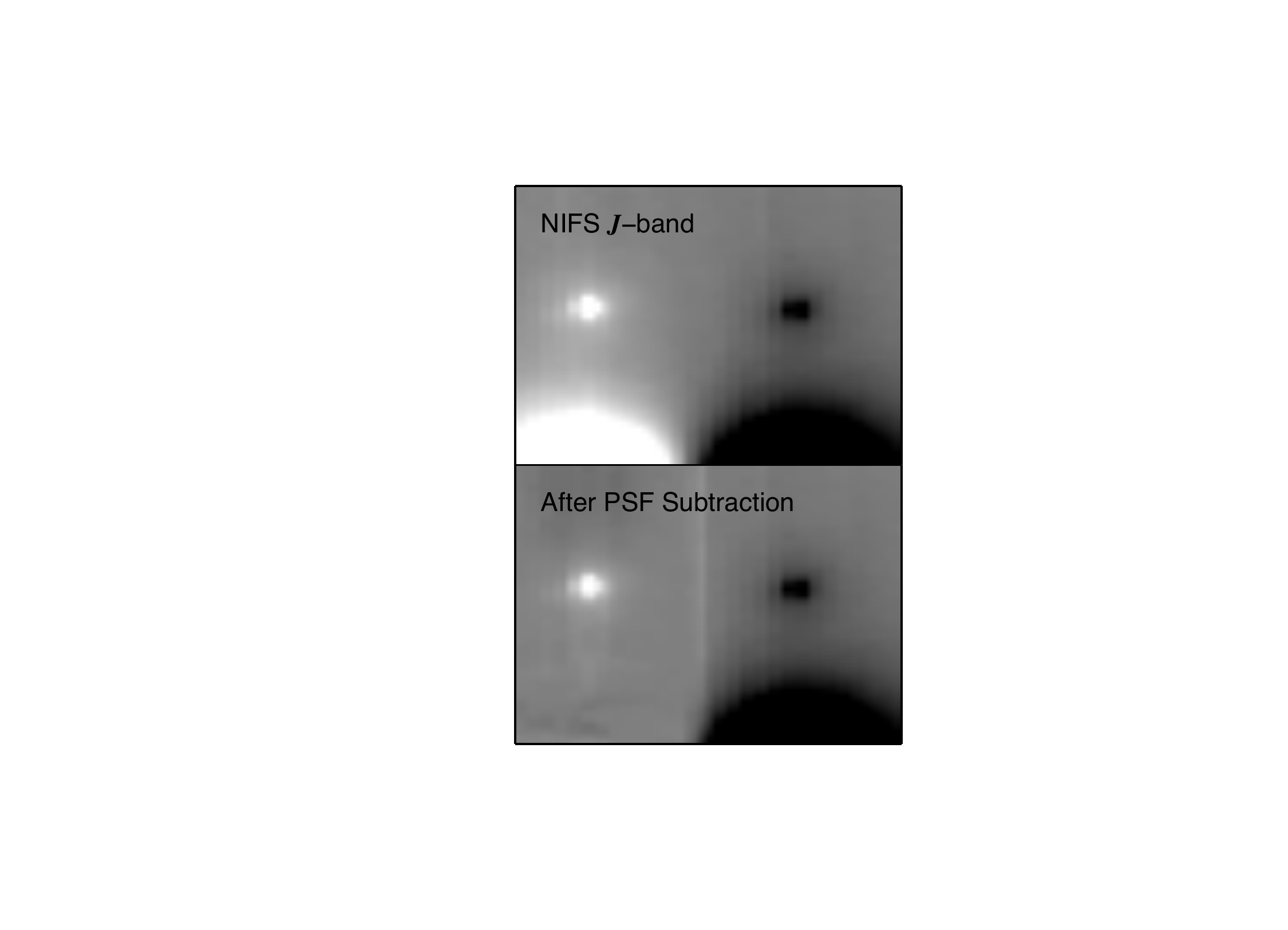}}
  \vskip -1 in
  \caption{Gemini-North/NIFS collapsed image of ROXs 12 B before (top) and after (bottom) PSF subtraction of the host star
  using a Gaussian profile.  
  The positive and negative images are caused by pairwise subtraction of two consecutive frames nodded by 1$\farcs$5.  
  ROXs 12 A is positioned outside the field of view but uncorrected residual light is visible at the bottom of the array.     \label{fig:nifsimg} } 
\end{figure}

We acquired moderate-resolution ($R$$\equiv$$\lambda$/$\delta$$\lambda$$\approx$6000)
$J$-band spectroscopy of ROXs 12 B spanning 1.15--1.36 $\mu$m with the 
Near-Infrared Integral Field Spectrometer (NIFS; \citealt{McGregor:2003kv}) at Gemini-North 8.1 m telescope on
UT 2016 April 18 and 22 (Gemini Program ID: GN-2016A-Q-37).
The facility AO system ALTAIR (\citealt{Christou:2010hx}) provided diffraction-limited 
correction using the Gemini laser guide star system with
ROXs 12 as a bright ($R$ = 13.5~mag) tip tilt reference.
NIFS has a 3$''$$\times$3$''$ field of view with 0$\farcs$1$\times$0$\farcs$04 rectangular spaxels.
To better sample the PSF wing of the host star, we rotated the instrument so that the 
short side of the spaxels was oriented in the direction of the ROXs 12 A and B position angle (P.A.)
with the host star located immediately off the detector (see Figure \ref{fig:nifsimg}).
Our observations of ROXs 12 B were then carried out in an ABBA pattern by nodding 1$\farcs$5 orthogonal to the binary P.A.

We acquired a total of 40 min (eight exposures of 300 s each) with the $J$-band grating centered at 1.25 $\mu$m 
and ZJ filter on UT 2016 April 18 in queue mode.  On UT 2016 April 22 an additional 70 min (14 exposures of 300 s each) 
were taken with the same configuration.  On both nights the A0V standard HD 145127 was targeted at a similar airmass for telluric correction.
Our observations are summarized in Table \ref{tab:specobs}.

\begin{figure*}
  \vskip -.8 in
  \hskip 0.2 in
  \resizebox{6.8in}{!}{\includegraphics{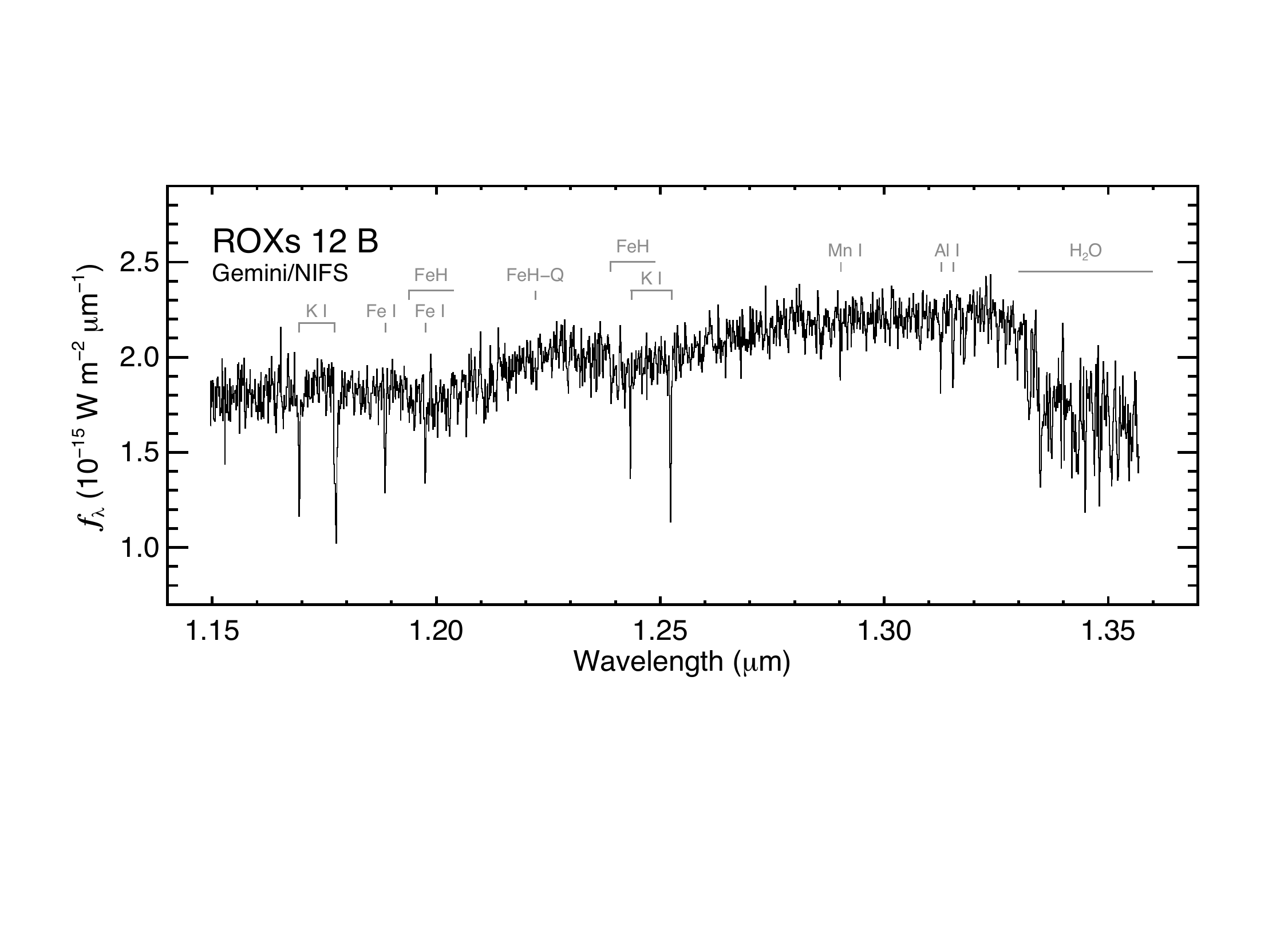}}
  \vskip -1.2 in
  \caption{Gemini-North/NIFS 1.15--1.36 $\mu$m spectrum of ROXs 12 B.  
  Strong atomic and molecular species are labeled in gray, including \ion{K}{1}, \ion{Fe}{1}, \ion{Mn}{1}, and \ion{Al}{1}.
  The resolving power is $\approx$6000 and the spectrum  has been corrected for reddening ($A_V$=1.8~mag)}.  \label{fig:nifsspec} 
\end{figure*}

Basic data reduction was carried out with NIFS reduction packages in IRAF which include flat fielding,
bad pixel interpolation, sky subtraction, and image rectification to data cubes.
ROXs 12 B is clearly visible in each cube but overlaps with the PSF wing from the host star.
To remove this low-level contaminating flux, we performed PSF subtraction in the same fashion
as in \citet{Bowler:2014dk} for NIFS data of ROXs 42B b.
For each column of each spectral channel, we masked out the companion and then fit and subtracted various
parameterized PSF models to the 2D wing of ROXs 12 A using the Levenberg-Marquardt least-squares curve fitting package 
\texttt{MPFIT} (\citealt{Markwardt:2009wq}).  
Gaussian, Lorentzian, and Moffat PSF models were tested to examine their influence on the extracted spectra.
All three profiles resulted in 
 similar spectra; ultimately we adopted the Gaussian model as this produced the lowest systematic over- or under-subtraction
in the region surrounding ROXs 12 B (Figure~\ref{fig:nifsimg}).
After PSF subtraction we extracted the spectra from each cube using aperture photometry 
and median-combined them after scaling individual spectra to their median values.
Telluric correction was  performed using
the \texttt{xtellcor\_general} routine in the Spextool data reduction package for IRTF/SpeX (\citealt{Vacca:2003wi}; \citealt{Cushing:2004bq}).
These steps were carried out separately for each night and the resulting spectra were subsequently combined 
by calculating the weighted mean and uncertainty of both spectra.
Note that standards were taken immediately before and after the science observations and the S/N levels were low for individual spectra,
so we performed telluric correction on the coadded spectra from each night instead of carrying this out for each separate exposure.
The final NIFS spectrum has an effective on-source integration time of 110 min and is shown in Figure~\ref{fig:nifsspec}
after having been dereddened by $A_V$=1.8 mag--- the extinction to ROXs 12 A measured by 
\citet{Rizzuto:2015bs} based on moderate-resolution optical spectroscopy--- following the extinction curve from \citet{Fitzpatrick:1985ih}.

% Figure 2

% Figure 3

% Figure 4

\begin{figure*}
  \vskip -.8 in
  \hskip 0.2 in
  \resizebox{6.8in}{!}{\includegraphics{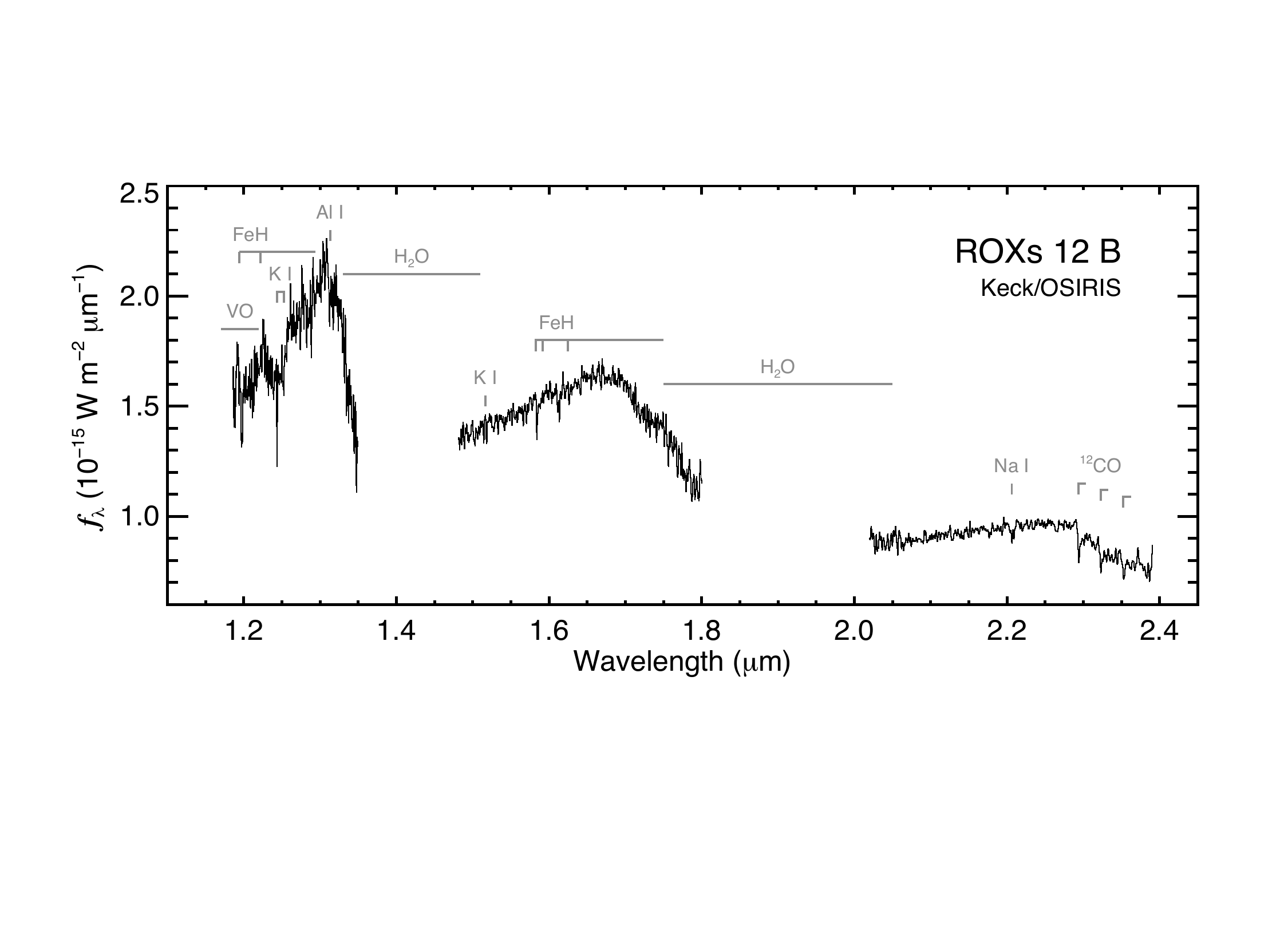}}
  \vskip -1.2 in
  \caption{Keck/OSIRIS 1.18--2.38 $\mu$m spectrum of ROXs 12 B.  Major atomic and molecular species are labeled in gray.
  Individual bandpasses have
  been flux-calibrated to photometry listed in Table~\ref{tab:properties}, Gaussian smoothed from a native resolving power of $R$=3800 to $R$=1000 
  to improve the S/N for visual purposes, and de-reddened by $A_V$=1.8~mag.
    \label{fig:ospec} } 
\end{figure*}

\subsection{Keck/OSIRIS Near-Infrared Spectroscopy of ROXs 12 B}{\label{sec:osiris}}

We targeted ROXs 12 B with the OH-Suppressing InfraRed Imaging Spectrograph (OSIRIS; \citealt{Larkin:2006jd}; \citealt{Mieda:2014dt})
using natural guide star adaptive optics (NSG AO; \citealt{Wizinowich:2013dz}) at the Keck~I telescope on 2016 May 22 UT.
Broad band filters were used with the 50 mas per spaxel plate scale, producing a 16$\times$64 spaxel (0$\farcs$8$\times$3$\farcs$2)
rectangular field of view with a spectral resolution of $R$$\approx$3800.
The sky was clear with seeing between 0$\farcs$5--0$\farcs$8 throughout the night.
To minimize contamination from the host star we oriented the long axis of the array to be orthogonal to the binary position angle.
Our observations consisted of nodded ABBA sequences with about 1$''$ offsets for pairwise sky subtraction.
A total of 50 min of integration time was acquired in $Kbb$ band (10 300-s exposures), 60 min in $Hbb$ band (12 300-s exposures), 
and 90 min in $J$ band (18 300-s exposures).
Multiple A0V standards were targeted between filter changes at a similar airmass as the science observations (see Table \ref{tab:specobs}).

Basic data reduction was carried out using the OSIRIS data reduction pipeline.  Images were flat-fielded, corrected for bad pixels, and assembled into data
cubes using the latest rectification matrices provided by Keck Observatory.  We then extracted the spectra from the data cubes using aperture photometry
at each wavelength at the centroided location of the target in median-collapsed cubes.  
Spectra were median-combined after scaling them to their individual median levels.
Telluric correction was then carried out 
with the \texttt{xtellcor\_general} routine in Spextool (\citealt{Vacca:2003wi}; \citealt{Cushing:2004bq}).
Each spectral band was then flux calibrated using  photometry of ROXs 12 B (Figure~\ref{fig:ospec}), which 
was derived by transforming the 2MASS photometry of ROXs 12 A to the MKO system using
relations from \citet{Leggett:2006gg} and then using the relative photometry of ROXs 12 AB from \citet{Kraus:2014tl}.  
The resulting photometry is reported in Table~\ref{tab:properties}.

\subsection{Harlan J. Smith Telescope/IGRINS High-Resolution Near-Infrared Spectroscopy}{\label{sec:igrins}

We obtained high-resolution ($R$=45000) 1.45--2.45 $\mu$m spectra of 
ROXs 12 A and 2M1626--2527 on UT 2016 July 26 with the Immersion Grating Infrared Spectrometer 
(IGRINS; \citealt{Park:2014kn}; \citealt{Mace:2016ev}) at the McDonald Observatory's 2.7-m Harlan J. Smith Telescope to measure 
radial velocities for both components of this common proper motion pair in order to test whether they share common space velocities.
Six exposures of 300 s each were acquired while nodding in an ABBA pattern along the slit,
totaling 30 min of integration for each target.
The A0V star HD 155379 was observed on the same night.
Spectra were extracted and reduced using version 2.1 of the IGRINS reduction pipeline\footnote{https://github.com/igrins/plp}
following the description in \citet{Bowler:2017hq}.
The S/N per pixel of both spectra range from 80--100.
In summary, after bias subtraction and flat fielding, the spectra were optimally 
extracted and cross correlated with over one hundred other IGRINS spectra of comparable spectral type
taken over the past two years to measure radial velocities.
The final radial velocities for ROXs 12 A and 2M1626--2527 are --6.09 $\pm$ 0.16 km s$^{-1}$ 
and --6.03 $\pm$ 0.16 km s$^{-1}$, respectively,
where the uncertainties are dominated by the transformation to the external reference frame
(as opposed to relative uncertainties, which are 0.04 km s$^{-1}$).

\subsection{Mayall/Ritchey-Chretien Spectrograph Optical Spectroscopy of 2M1626--2527}{\label{sec:goodman}}

We observed 2M1626--2527 with the Ritchey-Chretien Spectrograph (RC-Spec) using 
the T2KA CCD at the Kitt Peak National Observatory's 4-m Mayall telescope on UT 2014 May 21.
The 1$\farcs$5$\times$98$''$ slit was used with the BL420 grating resulting in a resolving power
of $R$$\approx$2600 spanning 6300--9200~\AA.
The slit was oriented in a fixed North-South direction throughout the night. 
We targeted 2M1626--2527 near transit at an airmass of 1.86 to minimize effects from 
differential atmospheric refraction.  
However, because the slit was not oriented exactly at parallactic angle, some slit losses may occur
which may have affected the slope of our RC-Spec spectrum.  These data are therefore useful for
spectral classification but not for detailed reddening measurements.  
A single exposure was acquired with the GG495 filter and a total integration time of 1200~s.

The raw data was first bias-subtracted and flat-fielded to remove pixel-to-pixel sensitivity variations. 
Night sky lines were removed using median sky values in the spatial direction on either side of the spectrum.
We then extracted the spectrum of the science target by summing flux in the spatial direction.
The overall throughput response was then corrected using the spectrophotometric standard HZ 44.
Finally, wavelength calibration was achieved using HeNeAr lamp observations taken throughout the night.
The optical spectrum of 2M1626--2527 is discussed in more detail in Section~\ref{sec:disk}.

\subsection{IRTF/SpeX Moderate-Resolution Near-Infrared Spectroscopy of 2M1626--2527}{\label{sec:spex}}

2M1626--2527 was targeted with the InfraRed Telescope Facility's SpeX instrument (\citealt{Rayner:2003vo}) 
in short cross-dispersed (SXD) mode on UT 2011 April 29.
We used the 0$\farcs$3 slit rotated to the parallactic angle, which yielded a mean resolving power of $\approx$2000
across the 0.8--2.4 $\mu$m spectrum.
Four nodded pairs were obtained with 60 s per exposure in an ABBA pattern.
The A0V standard HD 144925 was observed prior to our science target.
Standard data reduction including spectral extraction, wavelength calibration, and telluric correction was carried out 
using the Spextool package (\citealt{Vacca:2003wi}; \citealt{Cushing:2004bq}).
The near-infrared spectrum of 2M1626--2527 is discussed in more detail in Section~\ref{sec:disk}.

\subsection{Keck/HIRES High-Resolution Optical Spectroscopy}{\label{sec:hires}}

We obtained high-resolution ($R$$\approx$48000) optical spectra of ROXs 12 A and 2M1626--2527
with HIRES (\citealt{Vogt:1994tb}) at Keck Observatory on UT 2011 June 28 and UT 2015 May 3, respectively.
Both spectra were optimized for long wavelengths with the red cross disperser.  A single 300 s exposure of ROXs 12 A was taken with the C1 decker using the \emph{KV418} order blocking filter, resulting in a slit size of 7$\farcs$0$\times$0$\farcs$861 
on the sky and a wavelength range of 4310--8770 \AA.
For 2M1626--2527, we obtained one 600 s exposure spanning a wavelength range of 4800--9220 \AA \
using the C1 decker, the \emph{GG475} order blocking filter,
and a slit size of 7$\farcs$0$\times$0$\farcs$861.

Basic data reduction and spectral extraction were carried out following the description in \citet{Kraus:2011ju}.
The extraction pipeline MAKEE was used for bias subtraction, flat fielding, cosmic ray rejection, spectral extraction, and 
wavelength calibration.
Zero-point corrections to the wavelength solution were derived by cross correlating the telluric band
of the O7V star S Mon.
Radial velocities and projected rotational velocities ($v$sin$i_*$) were then measured for both targets using 
 early-M dwarf RV standards from \citet{Chubak:2012tv} after removing 
orders with strong telluric features or low S/N.
A broadening function (\citealt{Rucinski:1999uu}) was measured relative to these standards to find the 
absolute RV and line broadening, which was then translated into projected rotational velocities using
rotationally broadened template spectra to empirically correlate line broadening and 
projected rotational velocities following \citet{Kraus:2017bg}.
Our final RV values from HIRES are --5.67 $\pm$ 0.12 km $^{-1}$ for ROXs 12 
and --6.97 $\pm$ 0.15 km s$^{-1}$ for 2M1626--2527, comparable to the radial velocities
we found from our IGIRNS spectra.
The slight differences ($<$1 km s$^{-1}$) between the HIRES and IGRINS RVs
are likely caused by jitter in these young and active stars.  
Our $v$sin$i_*$ measurements from this analysis are 8.2 $\pm$ 0.7 km s$^{-1}$ and 4.5 $\pm$ 1.0 km s$^{-1}$
for ROXs 12 and 2M1626--2527, respectively.

% Figure 5

\begin{figure}
  \vskip -1.1 in
  \hskip -1.3 in
  \resizebox{5.9in}{!}{\includegraphics{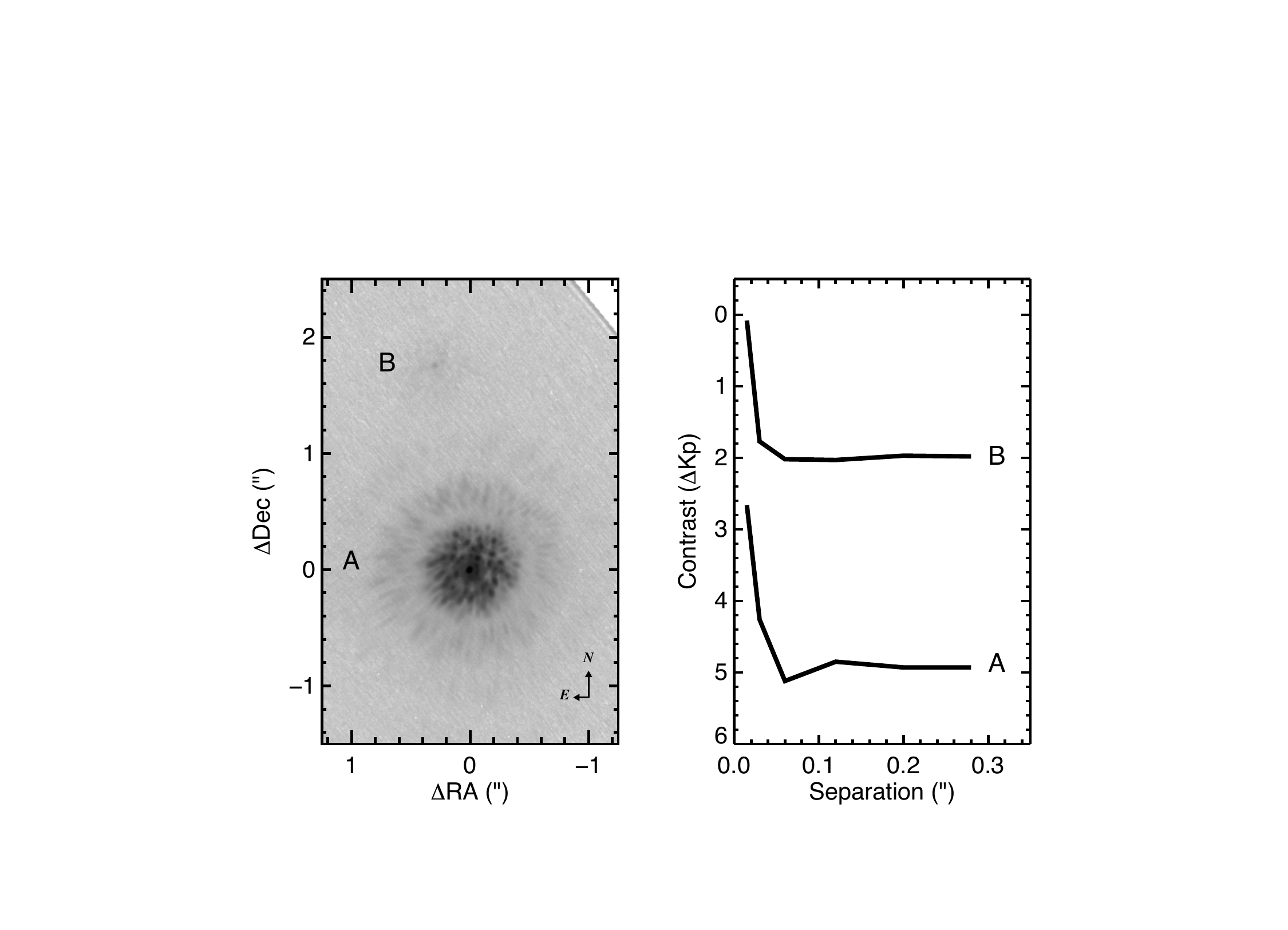}}
  \vskip -0.4 in
  \caption{Adaptive optics aperture masking interferometry of ROXs 12 A and B with the NIRC2 9-hole mask at Keck Observatory.  
  \emph{Left:} Interferograms of ROXs 12 A and B in this $Kp$-band image 
  offer a way to search for additional companions below the telescope diffraction limit.  
  \emph{Right:}  99\% contrast curves for ROXs 12 B (this work) and the host star (see \citealt{Cheetham:2015es}).   \label{fig:ami} } 
  \vskip 0.2 in
\end{figure}

\begin{deluxetable*}{lcccc}
\renewcommand\arraystretch{0.9}
\tabletypesize{\small}
\tablecaption{Physical, Kinematic, and Photometric Properties of the ROXs 12 Triple System {\label{tab:properties}}}
\tablehead{
   \colhead{Property} & \colhead{ROXs 12 A}  &  \colhead{ROXs 12 B}  & \colhead{2M1626--2527}   & \colhead{Ref} 
        }       
\startdata
2MASS ID  &  J16262803--2526477  &  $\cdots$  &  J16262774--2527247  & 1 \\ 
$B$ (mag) & 16.172 $\pm$ 0.141  &  $\cdots$  &  17.1 $\pm$ 0.7  &  2  \\
$V$ (mag) & 14.346 $\pm$ 0.066  &  $\cdots$  & 15.8 $\pm$ 0.3  &  2  \\
$R$ (mag) & 13.5  &  $\cdots$  &  15.3  &  3  \\
$r'$ (mag) & 13.5  &  $\cdots$  &  15.8   &  4  \\
$I$ (mag) & 11.870 $\pm$ 0.04  &  $\cdots$  & 12.91 $\pm$ 03  &  5  \\
%$J$ (mag) & 10.137 $\pm$ 0.07  &  $\pm$  &  $\pm$   &  DENIS  \\
%$K$ (mag) & 8.899 $\pm$ 0.08  &  $\pm$  &  $\pm$   &  DENIS  \\
$J_{2MASS}$ (mag) & 10.282 $\pm$ 0.024  & 15.82 $\pm$ 0.03\tablenotemark{a}  & 11.021 $\pm$ 0.024  &  1, 6\\
$H_{2MASS}$ (mag) & 9.386  $\pm$ 0.026  & 14.83 $\pm$ 0.03\tablenotemark{a} &  9.930 $\pm$ 0.026 &  1, 6  \\
$K_{s, 2MASS}$ (mag) &  9.099 $\pm$ 0.025 & 14.14 $\pm$ 0.03\tablenotemark{a} &  9.211 $\pm$ 0.025  &  1, 6 \\
$L'$  (mag) & [9.1 $\pm$ 0.1]\tablenotemark{b} & 13.2 $\pm$ 0.1  & [9.2 $\pm$ 0.1]\tablenotemark{b}  &  6 \\
$W1$  (mag) & 8.805 $\pm$ 0.023  &  $\cdots$ & 8.360 $\pm$ 0.023  &  7 \\
$W2$  (mag) &  8.712 $\pm$ 0.021  & $\cdots$ & 7.831 $\pm$ 0.021  &  7 \\
$W3$  (mag) &  8.393 $\pm$ 0.042  & $\cdots$ & 5.983 $\pm$ 0.017  &  7 \\
$W4$  (mag)  &    6.558 $\pm$ 0.089  & $\cdots$ & 3.784 $\pm$ 0.025  &  7 \\
$\mu_{\alpha}$cos$\delta$ (mas yr$^{-1}$) &  --12.7 $\pm$ 2.3  & $\cdots$ &  --11.9 $\pm$ 2.3  &  8  \\
$\mu_{\delta}$  (mas yr$^{-1}$)  & --29.0 $\pm$ 2.3  & $\cdots$ & --30.5 $\pm$ 2.3  &  8  \\
$v_\mathrm{rad}$  (km s$^{-1}$)\tablenotemark{c} &  --6.09 $\pm$ 0.16  & $\cdots$ &    --6.03 $\pm$ 0.16  &  9  \\
$v_\mathrm{rad}$  (km s$^{-1}$)\tablenotemark{d} &  --5.67 $\pm$ 0.12  & $\cdots$ &    --6.97 $\pm$ 0.15  &  9  \\
$v\sin i$  (km s$^{-1}$)\tablenotemark{d}  &  8.2 $\pm$ 0.7  & $\cdots$ &  4.5 $\pm$ 1.0  &  9 \\
log($L_\mathrm{bol}$/$L_{\odot}$)\tablenotemark{e}  & --0.57 $\pm$ 0.06  &  --2.87 $\pm$ 0.06  &  --0.81 $\pm$ 0.06  &  9  \\
Mass ($M_{\odot}$)  &  0.65$^{+0.05}_{-0.09}$  & 0.0167 $\pm$ 0.0014 & 0.5$^{+0.1}_{-0.1}$ &  9  \\
$T_\mathrm{eff}$ (K)  &  3900 $\pm$ 100 & 3100$^{+400}_{-500}$  &   3700 $\pm$ 150  & 9  \\
$P_\mathrm{rot}$ (d)  &  9.1 $\pm$ 0.4 & $\cdots$  &   3.30 $\pm$ 0.05  & 9  \\
Age (Myr) &  6$^{+4}_{-2}$  &  6$^{+4}_{-2}$  & 8$^{+7}_{-4}$ Myr  &  9  \\
Spectral Type &  M0 $\pm$ 0.5  &  L0 $\pm$ 2  &  M1 $\pm$ 1  &  9, 10 \\
$i_*$ ($^{\circ}$)\tablenotemark{f}  & 77$^{+7}_{-9}$  & $\cdots$  &  17$^{+5}_{-4}$  &  9 \\
$R_*$ ($R_{\odot}$)\tablenotemark{g} & 1.14 $\pm$ 0.07  & $\cdots$  &  0.96 $\pm$ 0.08  &  9  \\
\enddata
\tablenotetext{a}{$J$ and $H$ band photometry for ROXs 12 B is on the MKO filter system.  The $Ks$ band photometry assumes $K_s$$\approx$$K'$
to convert the contrast measurement from \citet{Kraus:2014tl} to an apparent magnitude.}
\tablenotetext{b}{Estimated $L'$-band magnitudes assuming $K$--$L'$=0.0 $\pm$ 0.1~mag for M0 and M1 spectral types (\citealt{Golimowski:2004en}).}
\tablenotetext{c}{From our IGRINS high-resolution near-infrared spectra.}
\tablenotetext{d}{From our HIRES high-resolution optical spectra.}
\tablenotetext{e}{Assumes a distance of 137 $\pm$ 10 pc.}
\tablenotetext{f}{Line-of-sight stellar inclination.}
\tablenotetext{g}{Radius from bolometric luminosity and effective temperature.}
\tablerefs{(1) 2MASS (\citealt{Cutri:2003tp}); (2) APASS DR9 (\citealt{Henden:2016uu}); (3) USNO-B1.0 (\citealt{Monet:2003bw});
(4) CMC15 (\citealt{Muinos:2014ew}); (5) DENIS; (6) \citet{Kraus:2014tl}; (7) AllWISE Data Release (\citealt{Cutri:2014wx}); 
(8) HSOY (\citealt{Altmann:2017hw}), which includes data from $Gaia$ DR1 (\citealt{GaiaCollaboration:2016gd}); (9) this work; (10)  \citet{Rizzuto:2015bs}.
}
\end{deluxetable*}

\subsection{K2 Time Series Photometry}{\label{sec:k2}}

ROXs 12 (EPIC 203640875) and 2MASS J16262774--2527247 (EPIC 203637940) 
were both observed with $K2$ (\citealt{Howell:2014vua}), 
the extended mission of the $Kepler$ spacecraft, during Campaign 2 between 
2014 August 23 and 2014 November 10 (GO 2052, PI: K. Covey; GO 2063, PI: A. Kraus).
The raw long cadence (30-min) photometry for ROXs 12 were corrected for 
systematic features caused by pixel-to-pixel drift and 
spacecraft thruster firings using the reduction pipeline described in \citet{Vanderburg:2014bia}.
For 2MASS J16262774--2527247, we used the raw light curve instead of applying 
corrections because this star exhibits very large photometric variations--- up to $\sim$50\% in amplitude--- 
which are far larger than systematics present in the data.
Light curves for both targets are presented in Section~\ref{sec:k2periods}.

\subsection{Keck/NIRC2 Aperture Masking Interferometry}{\label{sec:ami}}

% Figure 6

\begin{figure}
  \vskip -.4 in
  \hskip -.6 in
  \resizebox{4.in}{!}{\includegraphics{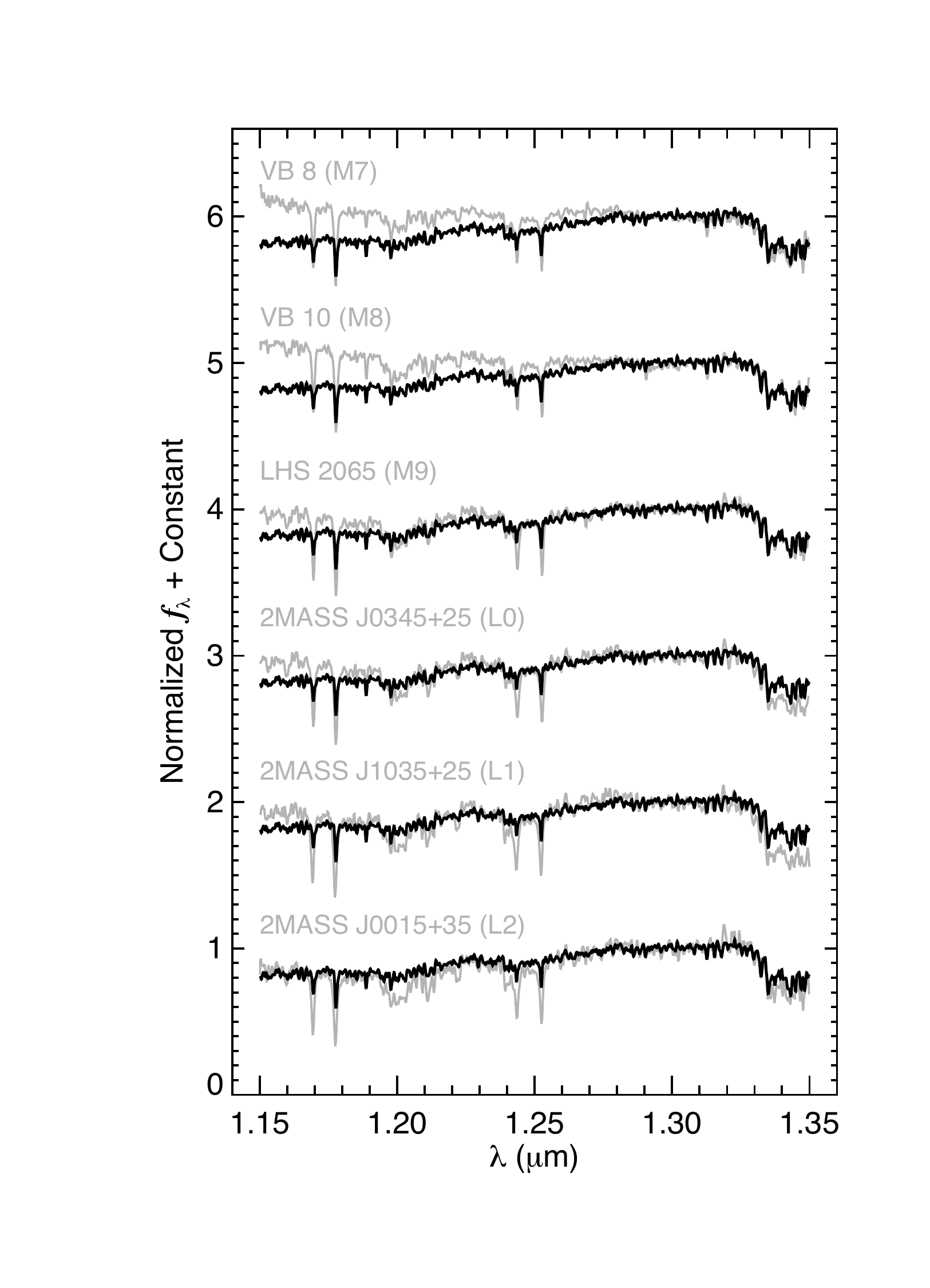}}
  \vskip -.5 in
  \caption{Comparison of our NIFS $J$-band spectrum of ROXs 12 B (black) to M7--L2 objects from BDSS (gray; \citealt{McLean:2003hx}; \citealt{McGovern:2004cc}).  
  Our spectrum has been Gaussian-smoothed to match the resolving power of the BDSS spectra ($R$$\approx$2000)
  and dereddened by $A_V$=1.8~mag to account for extinction.   \label{fig:bdss} } 
\end{figure}

ROXs 12 was observed on UT 2011 April 24 using 
NIRC2's 9-hole aperture mask together with natural guide star adaptive optics at Keck Observatory 
as part of the multiplicity survey of Ophiuchus members by \citet{Cheetham:2015es}.
The observations consist of four 20 s images (5 s $\times$ 4 coadds) in the $Kp$ filter using NIRC2's narrow camera mode.
After basic reduction (bias subtraction, flat fielding, and bad pixel correction),  
we re-examined the images and found that the interferogram from ROXs 12 B is clearly visible in all four frames.
This serendipitous detection opens the opportunity to search for a potential binary companion to ROXS 12 B at the
Keck diffraction limit--- analogous to brown dwarf-brown dwarf binary companions to stars like 
HD 130948 BC (\citealt{Potter:2002ie}) and $\epsilon$ Indi Bab (\citealt{Scholz:2003go}; \citealt{Mccaughrean:2004ey})--- 
albeit at modest flux ratios because of flux loss inherent to this technique and
ROXs 12 B's intrinsic faintness.

Figure \ref{fig:ami} displays the coadded image of ROXs 12 AB system after applying the
distortion solution and north orientation from \citet{Yelda:2010ig}.
Using the aperture masking pipeline described in \citet{Kraus:2008bh}, 
we did not find evidence for an additional companion to ROXs 12 B down 
to angular scales of about 30~mas.  The corresponding 99\%-level contrast curve in $Kp$ band over which we can exclude companions
is \{0.08, 1.77, 2.02, 2.03, 1.97, 1.98\} mag at \{15, 30, 60, 120, 200, 280\} mas.
This corresponds to mass limits of $\approx$10~\Mjup \ at about 0$\farcs$1 based on the evolutionary models of \citet{Baraffe:2015fwa} 
for an age of $\approx$6 Myr (see Section~\ref{sec:lum}).
It therefore appears that ROXs 12 B is single down to physical separations of $\approx$4~AU, which is well within its
Hill radius of about 50~AU.
Contrast limits for the host star from these same data are about 3 magnitudes deeper; 
these are shown in Figure~\ref{fig:ami} and presented in \citet{Cheetham:2015es}.

\section{Results}{\label{sec:results1}}

% Figure 7

\begin{figure*}
  \vskip -1.1 in
  \hskip -.55 in
  \resizebox{9.in}{!}{\includegraphics{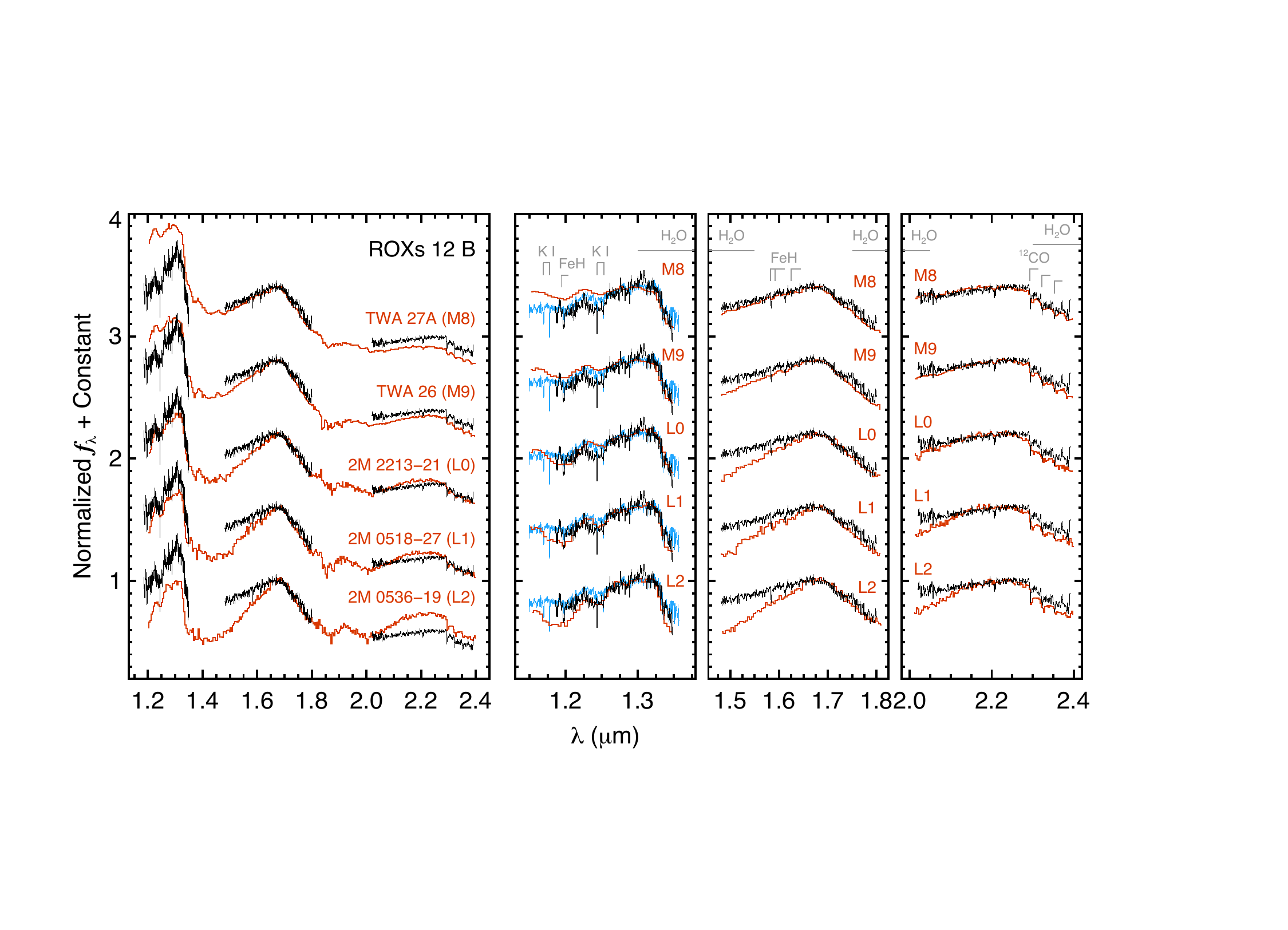}}
  \vskip -1.2 in
  \caption{OSIRIS (black) and NIFS (blue) spectra of ROXs 12 B compared to very low gravity standards from \citet{Allers:2013hk}.   
  ROXs 12 B shows conflicting properties: the 1.2--2.4~$\mu$m most resembles the L0 template, but the individual bandpasses
  are closer to L1--L2 in $J$ band and M8--M9 for $H$ and $K$ bands.  In particular, the blue side of the $H$ band slope
  is shallower than typical L dwarfs.  We adopt L0 $\pm$ 2 because of these discrepancies.  The OSIRIS and NIFS spectra
  have been de-reddened by $A_V$=1.8~mag to account for extinction.
  \label{fig:vlgcomp} } 
\end{figure*}

\begin{deluxetable}{lcc}
\renewcommand\arraystretch{0.9}
\tabletypesize{\small}
\tablewidth{0pt}
\tablecolumns{3}
\tablecaption{ROXs 12 B NIFS $J$-Band Equivalent Widths \label{tab:roxs12bew}}
\tablehead{
  \colhead{$\lambda_0$\tablenotemark{a}} & \colhead{Line}  &   \colhead{EW}  \\
     \colhead{($\mu$m)} & \colhead{ID}  &   \colhead{(\AA)}  
        }       
\startdata
1.1693 &  \ion{K}{1}      &   1.10 $\pm$  0.09 \\
1.1773  &  \ion{K}{1}     &    2.60 $\pm$ 0.10 \\
1.1887  &  \ion{Fe}{1}   &   0.89  $\pm$ 0.04 \\
1.1976  &  \ion{Fe}{1}   &   0.73  $\pm$ 0.05 \\
1.2436  &  \ion{K}{1}     &   0.88 $\pm$  0.04 \\
1.2526  &  \ion{K}{1}     &  1.26 $\pm$ 0.05  \\
1.2903   &  \ion{Mn}{1}  &  0.42 $\pm$ 0.12 \\
1.3127  &   \ion{Al}{1}   &  0.54 $\pm$  0.06 \\
1.3154  &  \ion{Al}{1}    &  0.47 $\pm$  0.03 \\
\enddata
\tablenotetext{a}{Nominal air wavelength.}
\end{deluxetable}

\begin{deluxetable*}{lcccccc}
\renewcommand\arraystretch{0.9}
\tabletypesize{\small}
\tablewidth{0pt}
\tablecolumns{5}
\tablecaption{\ion{K}{1} Equivalent Widths of Young Planetary-Mass Companions \label{tab:pmcew}}
\tablehead{
 \colhead{Object}  &  \colhead{SpT}  &   \colhead{\ion{K}{1} 1.169 $\mu$m} & \colhead{\ion{K}{1} 1.177 $\mu$m}  &   \colhead{\ion{K}{1} 1.243 $\mu$m}  &   \colhead{\ion{K}{1} 1.253 $\mu$m} & \colhead{Reference} \\
   & &   \colhead{EW (\AA)} &  \colhead{EW (\AA)} &    \colhead{EW (\AA)} &  \colhead{EW (\AA)}  & 
        }
\startdata
GSC 6214-210 B & M9.5 $\pm$ 1  &  $\cdots$  &  $\cdots$  &  3.0 $\pm$  0.5  &  3.2 $\pm$ 0.2  &  1 \\
ROXs 12 B & L0 $\pm$ 2  & 1.10 $\pm$ 0.09 & 2.60 $\pm$ 0.10 &  0.88 $\pm$  0.04  &  1.26 $\pm$ 0.05  &  2 \\
ROXs42B b &  L1 $\pm$ 1 & 2.8 $\pm$ 0.5  &  3.12 $\pm$ 0.19  &  1.9 $\pm$  0.3  &  1.76 $\pm$ 0.16 & 1 \\
2M0441+2301 Bb & L1 $\pm$ 1  &   $\cdots$  &  $\cdots$  &  5.4 $\pm$  0.6  &  4.0 $\pm$ 0.6  & 3 \\
1RXS J1609--2105 B & L2 $\pm$ 1  & $\cdots$   &  $\cdots$  &  5.33 $\pm$  1.0  &  3.8 $\pm$ 0.4 & 4, 5, 6 \\
\enddata
\tablerefs{(1) \citet{Bowler:2014dk}; (2) This work; (3) \citet{Bowler:2015en}; (4) \citealt{Lafreniere:2010cp}; (5) \citealt{Manjavacas:2014hu}; (6) \citealt{Wu:2015cz}.}
\end{deluxetable*}

% Figure 8

\begin{figure}
  \vskip -.4 in
  \hskip -.3 in
  \resizebox{3.8in}{!}{\includegraphics{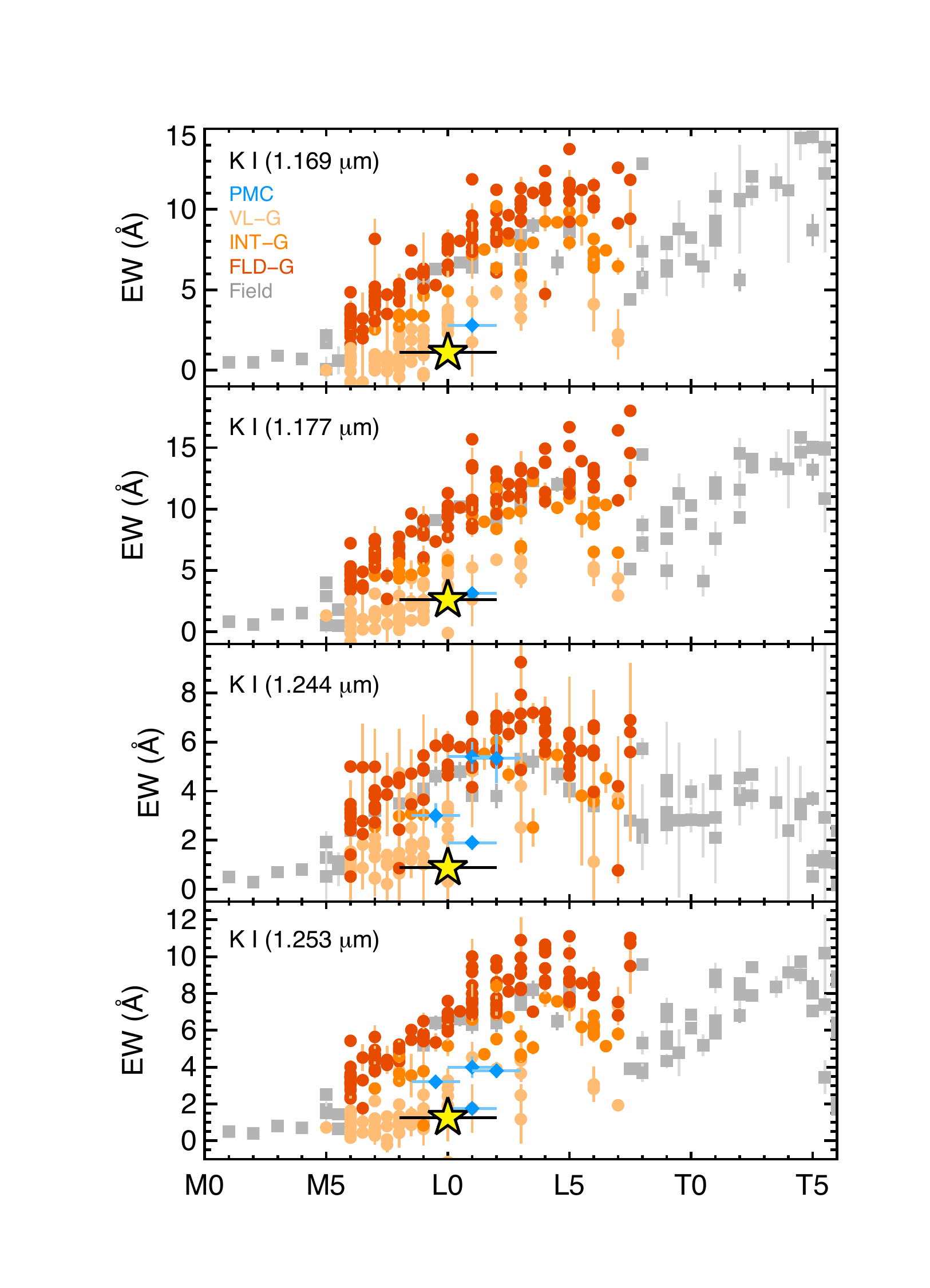}}
  \vskip -.3 in
  \caption{Equivalent widths of gravity-sensitive potassium features as a function of spectral type.
  Field objects without formal gravity classifications (gray squares) are from \citet{Cushing:2005ed} and \citet{Martin:2017jf}; 
  data for very low gravity (``VL-G''), intermediate-gravity (``INT-G''), and field gravity (``FLD-G'') are from \citet{Allers:2013hk} and \citet{Martin:2017jf}; 
  and young planetary-mass companions (PMCs) are from Table~\ref{tab:pmcew}.  
  The yellow star denotes the position of ROXs 12 B. \label{fig:k1ew} } 
\end{figure}

\subsection{Spectral Properties of ROXs 12 B}{\label{sec:roxs12bempirical}}

Our moderate-resolution spectra of ROXs 12 B from Gemini-N/NIFS ($J$ band) and 
Keck/OSIRIS ($J$, $H$, and $K$ bands) are shown in 
Figures~\ref{fig:nifsspec} and \ref{fig:ospec}.  
The NIFS spectrum shows prominent atomic and molecular absorption features from 
\ion{K}{1}, \ion{Fe}{1}, \ion{Mn}{1}, \ion{Al}{1}, H$_2$O, and FeH, as well as hints of
broad VO absorption at $\approx$1.2~$\mu$m.  Notably absent is Pa$\beta$ emission
at 1.282 $\mu$m, a signature of active accretion from a circumsubstellar disk;  
such emission has been detected in several other young brown dwarf companions 
over the past few years (e.g., \citealt{Seifahrt:2007iq}; \citealt{Bowler:2011gw}).

We compare our NIFS spectrum to brown dwarfs from the Brown Dwarf Spectroscopic Survey
(BDSS; \citealt{McLean:2003hx}) in Figure~\ref{fig:bdss}.  
Relative to field objects, ROXs 12 B exhibits shallower \ion{K}{1} doublets at 1.169/1.178 $\mu$m and 1.244/1.253 $\mu$m
as well as diminished FeH molecular absorption features at $\approx$1.195--1.205 and 1.239 $\mu$m--- 
all signs of low surface gravity (e.g., \citealt{Gorlova:2003cs}; \citealt{Slesnick:2004jy}; \citealt{McGovern:2004cc}; \citealt{Allers:2013hk}).
The pseudocontinuum is a good match to the L2 template spectrum.
Equivalent widths of absorption lines are reported in Table~\ref{tab:roxs12bew}.

Our OSIRIS spectrum broadly resembles a late-M and early-L dwarf with
deep H$_2$O bands at $\approx$1.4 and $\approx$1.9 $\mu$m, FeH, $^{12}$CO, \ion{Na}{1}, and 
weak \ion{K}{1}.  The triangular $H$ band shape is typical of low-gravity brown dwarfs
and giant planets (e.g., \citealt{Lucas:2001ed}; \citealt{Kirkpatrick:2006hb}; \citealt{Allers:2007ja}),
offering independent evidence that the system is young, although note that 
the triangular $H$-band shape does not strictly track low surface gravity (\citealt{Allers:2013hk}).

In Figure \ref{fig:vlgcomp} we compare our OSIRIS and NIFS spectra to ``very low gravity'' (``VL-G'') templates from \citet{Allers:2013hk}.
The full 1.2--2.4~$\mu$m spectrum is most similar to the young L0 template, but individually each bandpass shows
inconsistent matches: the $J$ band resembles L1--L2 objects, the $H$ band is closest to the M8 template,
and the $K$ band appears similar to M8--M9 objects.  Altogether we adopt a near-infrared spectral type
of L0 $\pm$ 2 for ROXs 12 B.  Note that  the OSIRIS broadband filter 
bandpasses do not cover spectral regions used in the \citet{Allers:2013hk} classification system so we 
adopt visual-based classifications for ROXs 12 B.

Equivalent widths of the $J$-band potassium doublets from our NIFS spectrum 
are shown relative to ultracool objects in Figure~\ref{fig:k1ew}.  
High-gravity field objects possess deep lines which trace out an upper envelope of equivalent widths,
whereas young objects exhibit weaker alkali features as a result of their low-pressure atmospheres.
Compared to low-mass stars and brown dwarfs spanning a range of ages and surface gravities,
ROXs 12 B has shallow potassium lines similar to very low-gravity objects 
from \citet{Allers:2013hk} and \citet{Martin:2017jf} as well as
young ($\lesssim$10~Myr) companions near and below the deuterium burning
limit, as measured from published spectra (Table~\ref{tab:pmcew}).

\subsection{Physical Properties of ROXs 12 B}{\label{sec:roxs12bphysical}}

We implement two methods to assess the effective temperature and
surface gravity of ROXs 12 B using atmospheric models.  The first approach
is based on maximum likelihood fitting of the BT-Settl models (``CIFIST2011bc'' version;
\citealt{Allard:2011jx}) following the general description in \citet{Bowler:2009hn}.
In summary, the grid of models are first Gaussian smoothed to the resolving power of
the OSIRIS spectrograph ($R$$\approx$3800) and resampled onto 
the same wavelength grid as the data.  A reduced chi squared value ($\chi^2_{\nu}$) 
is calculated for each synthetic spectrum in the grid spanning \Teff = 1100--4000~K ($\Delta$\Teff = 100 K)
and log~$g$ = 2.5--5.5 dex [cgs units throughout this work] ($\Delta$log~$g$ = 0.5 dex).  Similarly, a corresponding
radius is derived for each model by making use of the factor that scales the emergent model spectrum
to the observed flux-calibrated spectrum, as this quantity is also equal to the ratio of 
the object's radius to its distance, squared.  Here we adopt a distance of 137 $\pm$ 10 pc, where the mean value 
reflects the distance measured to the dark cloud 
Lynds 1688 by \citet{OrtizLeon:2017ce} and the uncertainty is reflects the 
uncertainty in the membership of this system to the Ophiuchus versus Upper Sco star-forming regions (see Section~\ref{sec:ophusco}).

%137.3 $\pm$ 1.2 pc measured to the dark cloud 
%Lynds 1688 by \citet{OrtizLeon:2017ce}, although we note that the Ophiuchus/Upper Scorpius star-forming region
%likely spans a much broader range of perhaps $\approx$10 pc.

In addition to fitting the entire 1.2--2.4~$\mu$m spectrum, we also fit the individual
$J$, $H$, and $K$ bands using this method.  Results are shown in Figure~\ref{fig:modfits}.
The models qualitatively match the data quite well but produce different quantitative results 
depending on the bandpass, with effective temperatures ranging from 1600 K to 2800 K.  
The bottom panel shows the entire 1.2--2.4~$\mu$m spectrum; the best model does a poor job of
reproducing the data by underestimating the flux at shorter wavelengths and over-predicting it in $K$ band.
  Despite the generally good agreement of the models with the individual bands, the inferred
gravities and effective temperatures disagree depending on the spectral region being 
considered.  
Two regions of local minima are clearly visible for the effective temperature, one
at $\approx$1500--1700~K and one at $\approx$2600--2900~K.  
For the entire spectrum, the best fitting model is visually a poor match to the data
despite formally producing the lowest $\chi^2_{\nu}$ value.  
Similarly, the inferred radii span 1.7--5.1~$R_\mathrm{Jup}$ 
depending on the spectral bandpass used in the fits.  
Only the $H$ and $K$ bands are consistent with expectations of $\approx$2~$R_\mathrm{Jup}$ 
from hot-start evolutionary models.  The implied masses based on surface gravity and radii are also unphysical,
ranging from 3.8~\Mjup \ for $H$ band to 3300~\Mjup \ for the entire spectral fit.
These discrepancies make it difficult to interpret the results; 
systematic errors in the models mean that the data are not standard deviates about the model
and therefore the $\chi^2_{\nu}$ statistic does not correspond to a maximum likelihood.
Instead, these fits are more sensitive to the overall shape of the pseudocontinuum rather than
individual absorption features which may be more useful for constraining
surface gravity and effective temperature.

% Figure 9

\begin{figure}
  \vskip -1 in
  \hskip -.2 in
  \resizebox{3.7in}{!}{\includegraphics{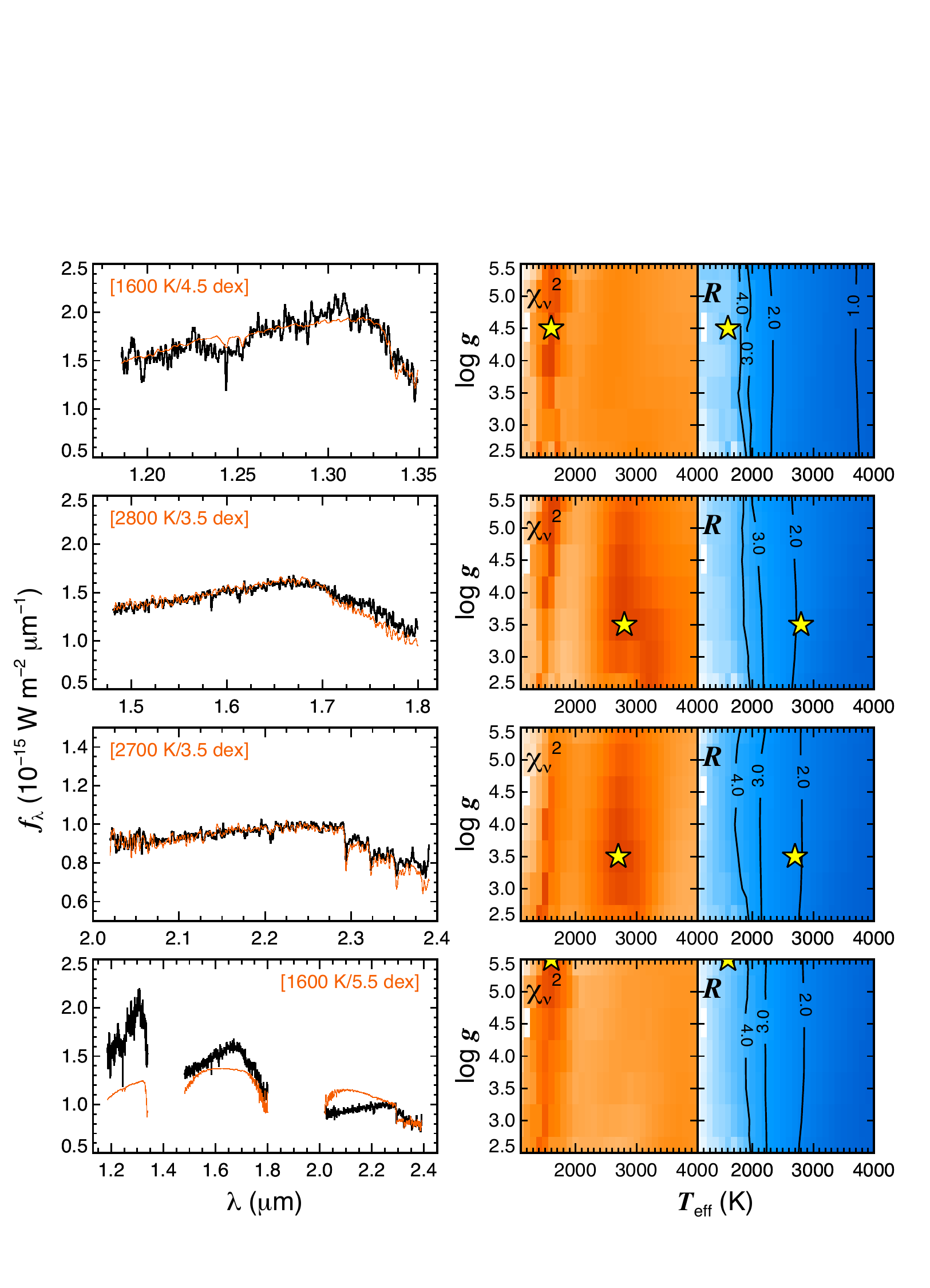}}
  \vskip -.3 in
  \caption{Best-fitting BT-Settl atmospheric models to our OSIRIS spectrum of ROXs 12 B.  The top three
  panels show results for individual $J$, $H$, and $K$ bandpasses.  
   Individual best-fitting models' effective temperature and surface gravity are listed in red.
  The right panels show the reduced $\chi^2$ maps and associated radius in $R_\mathrm{Jup}$ (using the scaling factor and distance to the system),
  with a star marking the global minimum.  
 \label{fig:modfits} } 
  \vskip -.1 in
\end{figure}

% Figure 10

\begin{figure}
  \vskip -.1 in
  \hskip -.2 in
  \resizebox{3.7in}{!}{\includegraphics{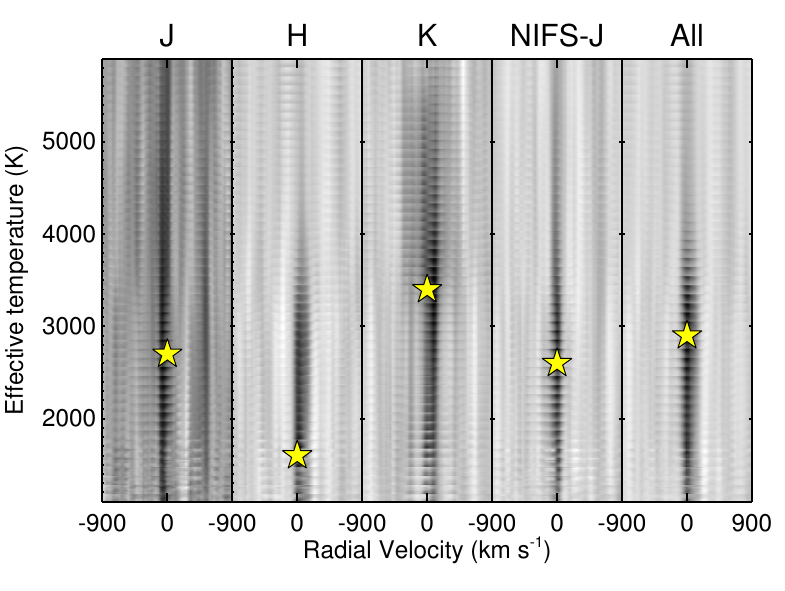}}
  \vskip -.1 in
  \caption{Cross correlation plots for BT-Settl synthetic spectra relative to our OSIRIS and NIFS spectra of ROXs 12 B.  
  Each row  shows the cross correlation function for that particular model in velocity space, with higher cross correlation values  
  represented as darker pixel values.  Each effective temperature is subdivided into rows with eight models spanning surface gravities  between 2.0--5.5~dex, which produces the horizontal structure visible in each panel. The right-most panel shows the cross correlation results after combining the individual bandpasses.  Yellow stars denote the maximum cross correlation defined as having the highest S/N value.
 \label{fig:ccfs} } 
  \vskip -.1 in
\end{figure}

For this reason we explored another approach based on maximizing the cross correlation between models and data. This method better incorporates information from atomic and molecular lines, which may be more sensitive to temperature and surface gravity, compared with the $\chi^2$ treatment. Cross correlation  encodes line positions and line ratios but is mostly insensitive to uncertainties in the calibration of the continuum as well as broad-band spectral variations (e.g., \citealt{Brogi:2012bk}).

The same set of models described above was convolved with a Gaussian kernel to a resolution of $R$ = 10,000. This is high enough to provide a high-enough resolution template for both the OSIRIS and the NIFS spectra, but low enough to avoid interpolation errors. Prior to cross-correlation, a high-pass filter 
was applied to both the data and the models to remove broad-band variations. The models were then Doppler-shifted to radial velocities between $-$900 and +900 km s$^{-1}$ (in steps of 30 km s$^{-1}$), linearly interpolated to the spectral range of the data, and cross-correlated with our observed spectra. Each spectral channel in the data was weighted by the inverse of its relative error to prevent noisy sections of the spectra from dominating the analysis.

The cross-correlation functions (CCFs) obtained with the full set of model spectra are shown in Figure~\ref{fig:ccfs}. Not only does the S/N vary between different bands, but also the peak of the cross-correlation seems to occur at different positions in the spectral sequence, confirming the previous observation that different bands could lead to different best-fitting parameters for the atmosphere.

To formally compute confidence intervals on the model parameters, we incorporated the values of the CCFs into a likelihood function $L$ following the approach of \citet{Zucker:2003it}. Since in this case we are combining four different bands, we adopt the following effective cross-correlation function:

\begin{equation}
\bar{C}^2 = 1 - \left\{\prod_i[1- C_i^2] \right\}^{1/4}
\end{equation}

\noindent where the product is performed over the cross-correlation function of the four available spectral bands. The effective cross-correlation function and the number of spectral channels $N$ are used to compute a log-likelihood function:

\begin{equation}
\log(L) = -\frac{N}{2}\log[1-\bar{C}^2]
\end{equation}

\noindent and the corresponding likelihood values $L$ are used to derive confidence intervals (CIs). 

The bottom-left panel in Figure~\ref{fig:ci_logL} shows the two-dimensional map of $\log(L)$ (blue shades) and the corresponding 1-, 2-, and 3-$\sigma$ CIs are labeled. The top-left and bottom-right panels show the projected probability distributions for $\log(g)$ and $T_\mathrm{eff}$, respectively. The resulting effective temperature of ROXs 12 B is $T_\mathrm{eff} = 3100^{+400}_{-500}$ K, and the bi-modality observed with the previous model comparison is no longer present. However, only an upper limit of $\log(g) < 4.0$ is obtained for the surface gravity. Although this result points to a low-gravity object, the likelihood also rises at high values of $\log(g)$.

% Figure 11

\begin{figure}
  \vskip -.1 in
  \hskip -.2 in
  \resizebox{3.7in}{!}{\includegraphics{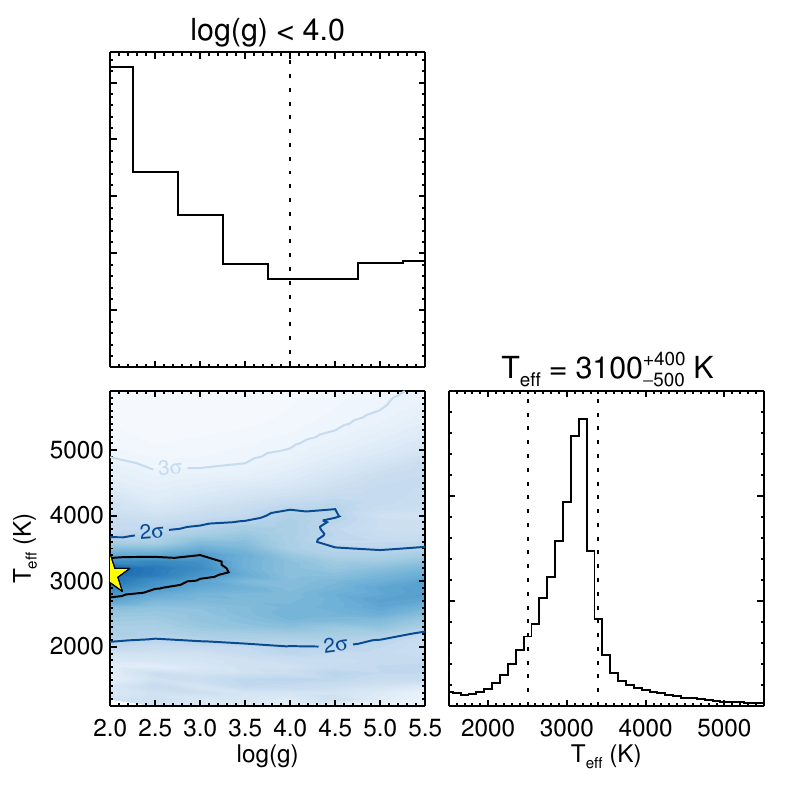}}
  \vskip -.1 in
  \caption{Cumulative cross correlation results for all of our moderate-resolution spectra of ROXs 12 B. 
  The bottom left panel shows the joint constraint on effective temperature and surface gravity, while the 
  upper panel and right panel display their projected one-dimensional distributions.  
  The constraints are generally broad, with a 1-$\sigma$ upper limit of 4.0 dex for the surface gravity and 
  an effective temperature of 3100$^{+400}_{-500}$ K.  However, as opposed to the $\chi^2$ approach 
 (e.g., Figure~\ref{fig:modfits}), the bimodality is removed and robust confidence intervals can be obtained.
 The yellow star denotes the peak of the joint cross correlation function.
 \label{fig:ci_logL} } 
  \vskip -.1 in
\end{figure}

The inferred effective temperature from our cross correlation analysis 
is warmer than expected for a young L0 object, although the distribution
is broad and encompasses the range of anticipated values.  For example,
using the effective temperature-spectral type relation for young objects
from \citet{Filippazzo:2015dv} yields 2260~$\pm$~60~K, assuming no
error on the spectral type.  If instead we adopt an uncertainty of 2 subtypes,
the median effective temperature of the resulting distribution is 2160~K
with a 1-$\sigma$ highest posterior density range\footnote{Also known as the minimum credible interval,
this is the region containing the highest values of a posterior distribution and may comprise
several intervals for multi-modal distributions.
}
 of 1810--2770~K.

We determine the luminosity of ROXs 12 B using our flux-calibrated OSIRIS spectrum together
with bolometric corrections using BT-Settl model synthetic spectra at shorter ($\lambda$$<$1.2~$\mu$m) and 
longer ($\lambda$$>$2.4~$\mu$m) wavelengths.  
The synthetic spectrum is scaled to the short- and long-wavelength endpoints of our spectrum
and then integrated to derive a bolometric flux.
Because the results from atmospheric model fits are either inconclusive (for the $\chi^2$ analysis)
or have large uncertainties (from the CCF analysis), we utilized the \Teff=2300~K model with $\log g$= 4.0 dex based on relations from  \citet{Filippazzo:2015dv} for young L0 objects--- 
although the resulting luminosities were only weakly sensitive to the input model.
The bolometric luminosity of ROXs 12 B is $\log L_\mathrm{bol}/L_{\odot}$ = --2.87 $\pm$ 0.06 dex,
which takes into account uncertainties in individual spectral measurements, the flux calibration
scale factor, and the estimated distance to the system.
In comparison, using a 3100~K model produces similar bolometric luminosity of 
--2.82 $\pm$ 0.06 dex.

Based on the bolometric luminosity of the companion and the system 
age (6$^{+3}_{-2}$ Myr; Section~\ref{sec:lum}), the inferred mass of ROXs 12 B is 17.5~$\pm$~1.5~\Mjup \
from \citet{Burrows:1997jq} hot-start evolutionary models.  
This mass is consistent with the value of 16 $\pm$ 4 \Mjup \ 
found by \citet{Kraus:2014tl}.

\subsection{2M1626--2527: A Wide Tertiary Companion}{\label{sec:binary}}

2M1626--2527s appear to constitute a wide tertiary to ROXs 12 AB 
separated by 37$''$, or  5100 AU at a distance of about 140 pc.  
The proper motions of this pair from HSOY (\citealt{Altmann:2017hw}) 
are consistent with one another:
$\mu_{\alpha}$cos$\delta$ = --12.7 $\pm$ 2.3 mas yr$^{-1}$ and $\mu_{\delta}$ = --29.0 $\pm$ 2.3  mas yr$^{-1}$ for ROXs 12,
and 
$\mu_{\alpha}$cos$\delta$ = --11.9 $\pm$ 2.3 mas yr$^{-1}$ and $\mu_{\delta}$ =  --30.5 $\pm$ 2.3  mas yr$^{-1}$ for 2M1626--2527.
Similarly, we measure  consistent radial velocities for both components from our IGRINS spectra:
--6.09 $\pm$ 0.16 km s$^{-1}$ for ROXs 12 and --6.03 $\pm$ 0.16 km s$^{-1}$ for 2M1626--2527.
This is far less than the  velocity dispersion of $\approx$ 1 km s$^{-1}$ for $\rho$ Ophiuchus 
cluster members (\citealt{Wilking:2015hn}), offering further evidence that these two stars are not simply members of the same
cluster, but are  likely gravitationally bound.  
The identical radial velocities also suggest that both stars are probably single.  
Indeed, deep adaptive optics imaging by \citet{Bryan:2016eo} did not
reveal any additional members of this system.
The kinematic, photometric, and physical properties of this triple system comprising
ROXs 12 A, ROXs 12 B, and 2M1626--2527 are listed in Table \ref{tab:properties}.

\subsection{Disk Properties and Accretion}{\label{sec:disk}}

% Figure 12

\begin{figure}
  \vskip -.35 in
  \hskip -0.3 in
  \resizebox{4.in}{!}{\includegraphics{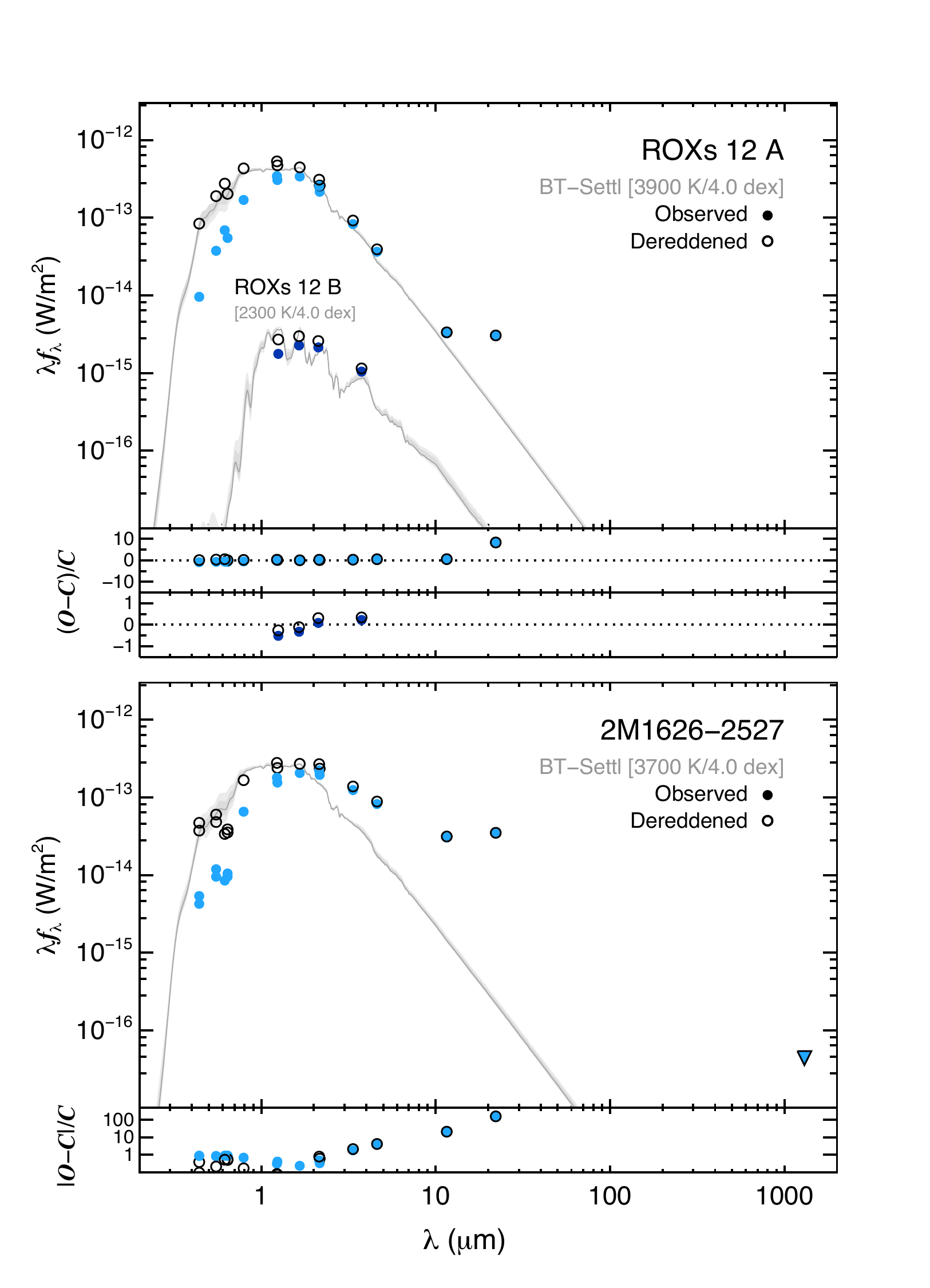}}
  \vskip -0.1 in
  \caption{Spectral energy distributions for ROXs 12 A, ROXs 12 B, and 2M1626--2527.  Raw photometry are shown
  as filled colored circles, while open symbols have been dereddened by $A_V$=1.8~mag.  BT-Settl models are shown
  in dark gray with the nominal effective temperature and gravity of each object, as well as $\pm$100~K ranges shaded in gray.
  Models have been scaled to the dereddened $H$-band flux densities.
        \label{fig:sedfig} } 
\end{figure}

Figure \ref{fig:sedfig} shows the spectral energy distributions of 
ROXs 12 A, ROXs 12 B, and 2M1626--2527 based on optical through 
infrared photometry in Table~\ref{tab:properties}.  
Flux density measurements for all three targets are dereddened by 
$A_V$=1.8 mag.

\begin{deluxetable}{lcc}
\renewcommand\arraystretch{0.9}
\tabletypesize{\small}
\tablewidth{0pt}
\tablecolumns{3}
\tablecaption{Emission Line Equivalent Widths \label{tab:ew}}
\tablehead{
  \colhead{$\lambda_0$\tablenotemark{a}} & \colhead{Line}  &   \colhead{EW}  \\
     \colhead{(\AA)} & \colhead{ID}  &   \colhead{(\AA)}  
        }       
\startdata
\cutinhead{ROXs 12 A: Keck/HIRES}
6562.8  &   H$\alpha$  &  --1.41 $\pm$ 0.02   \\
8498.0  &   \ion{Ca}{2}  &  --0.40 $\pm$ 0.01 \\
8542.1  &   \ion{Ca}{2}  & --0.60 $\pm$ 0.01 \\
8662.1  &   \ion{Ca}{2}   & --0.42 $\pm$ 0.02 \\
\cutinhead{2M1626--2527: Keck/HIRES}
4861.3  &  H$\beta$  & --26.2 $\pm$ 0.2  \\
4921.9  &  \ion{He}{1}  &   --0.53 $\pm$ 0.03 \\
5015.7  &  \ion{He}{1}  &    --0.51 $\pm$ 0.03 \\
5875.6  &  \ion{He}{1}   &  --2.54 $\pm$ 0.03  \\
6562.8  &   H$\alpha$  &  --53.17 $\pm$ 0.13   \\
6678.2  &   \ion{He}{1}  & --1.12 $\pm$ 0.02 \\
 8446\tablenotemark{b}   &  \ion{O}{1}   &   --0.47 $\pm$ 0.04 \\
8498.0  &   \ion{Ca}{2}  &  --1.1 $\pm$ 0.01 \\
8542.1  &   \ion{Ca}{2}  & --1.29 $\pm$ 0.02 \\
8662.1  &   \ion{Ca}{2}   & --0.82 $\pm$ 0.01 \\
\cutinhead{2M1626--2527: Mayall/RC-Spec}
     6300.3   &     [\ion{O}{1}]   &    --0.9 $\pm$ 0.3 \\
   6562.8   &    H$\alpha$    &    --46.5 $\pm$ 0.5 \\
  6678.2   &   \ion{He}{1}     &     --1.33 $\pm$ 0.12  \\
    7002\tablenotemark{b}   &      \ion{O}{1}    &     --0.4 $\pm$ 0.2 \\
    7065.2    &    \ion{He}{1}   &     --0.9 $\pm$ 0.3  \\
       8446\tablenotemark{b}    &     \ion{O}{1}   &   --1.09 $\pm$ 0.05 \\
    8498.0    &   \ion{Ca}{2}   &      --2.00 $\pm$ 0.05 \\ 
    8542.1    &   \ion{Ca}{2}    &     --1.91 $\pm$ 0.04 \\ 
     8662.1   &   \ion{Ca}{2}    &     --1.46 $\pm$ 0.04 \\ 
\cutinhead{2M1626--2527: IRTF/SpeX}
      8446\tablenotemark{b}   &          \ion{O}{1}   &      --0.7 $\pm$ 0.3 \\
    8498.0   &    \ion{Ca}{2}    &     --1.9 $\pm$ 0.3 \\
   8542.1    &   \ion{Ca}{2}   &      --1.6 $\pm$ 0.2 \\
  8662.1    &     \ion{Ca}{2}  &    --0.8 $\pm$ 0.2 \\
    9546.2   &  \ion{H}{1} Pa8 ($\epsilon$)  &     --0.9 $\pm$ 0.3 \\
 10049.8  &  \ion{H}{1} Pa7 ($\delta$)   &      --1.5 $\pm$ 0.2 \\
10830.3    & \ion{He}{1}   &     --1.1 $\pm$ 0.2 \\
  10938.2    & \ion{H}{1} Pa6 ($\gamma$)  &     --1.7 $\pm$ 0.2 \\
 128218.1     & Pa $\beta$   &   --1.85 $\pm$ 0.13 \\
21661.2   & Br $\gamma$   &       --1.5 $\pm$ 0.2 \\
\enddata
\tablenotetext{a}{Nominal air wavelength.}
\tablenotetext{b}{Lines are blended in our spectrum.}
\end{deluxetable}

Solar metallicity BT-Settl model atmospheres from \citet{Allard:2012fp} are overplotted 
in Figure \ref{fig:sedfig}.
Spectral types are converted to effective temperatures using 
empirical calibrations for pre-main sequence stars.
ROXs 12 A has a spectral type of M0.0 $\pm$ 0.5 (\citealt{Bouvier:1992ww}; \citealt{Rizzuto:2015bs}), which
corresponds to 
an effective temperature of 3900$^{+60}_{-90}$ K on the \citet{Herczeg:2014is} scale
and 3770$^{+100}_{-70}$ K on the \citet{Pecaut:2013ej} scale; we adopt an effective temperature
of 3900 $\pm$ 100 K for ROXs 12 A here.
Without a large sample of radius measurements for young stars, it is difficult to assess the
size of potential systematic errors in these spectral type-effective temperature scales.  However, we note that even systematic errors as large 
as 300~K for ultracool dwarfs (e.g., \citealt{Dupuy:2010ch})
do not substantially effect the broad results or conclusions regarding the luminosities, radii, or stellar inclinations throughout this paper.

The agreement of the model atmosphere with the dereddened photometry of ROXs 12 A is generally quite good spanning
the optical to mid-IR wavelengths, but the SED clearly shows a large excess at 22 $\mu$m,
presumably from disk emission originating from the host star ROXs 12 A or, more unlikely, a warm disk
surrounding the companion ROXs 12 B that is only prominent at 22 $\mu$m.
If this excess emission originates from ROXs 12 A and peaks at $\sim$22~$\mu$m then the dust temperature
is $\sim$230~K and is located at $\approx$0.7~AU following the heuristic relations from \citet{Wyatt:2008ht}.
If it originates from ROXs 12 B then it must be located much closer in at $\sim$0.05~AU.

% Figure 13

\begin{figure*}
  \vskip -.7 in
  \hskip 0.2 in
  \resizebox{7in}{!}{\includegraphics{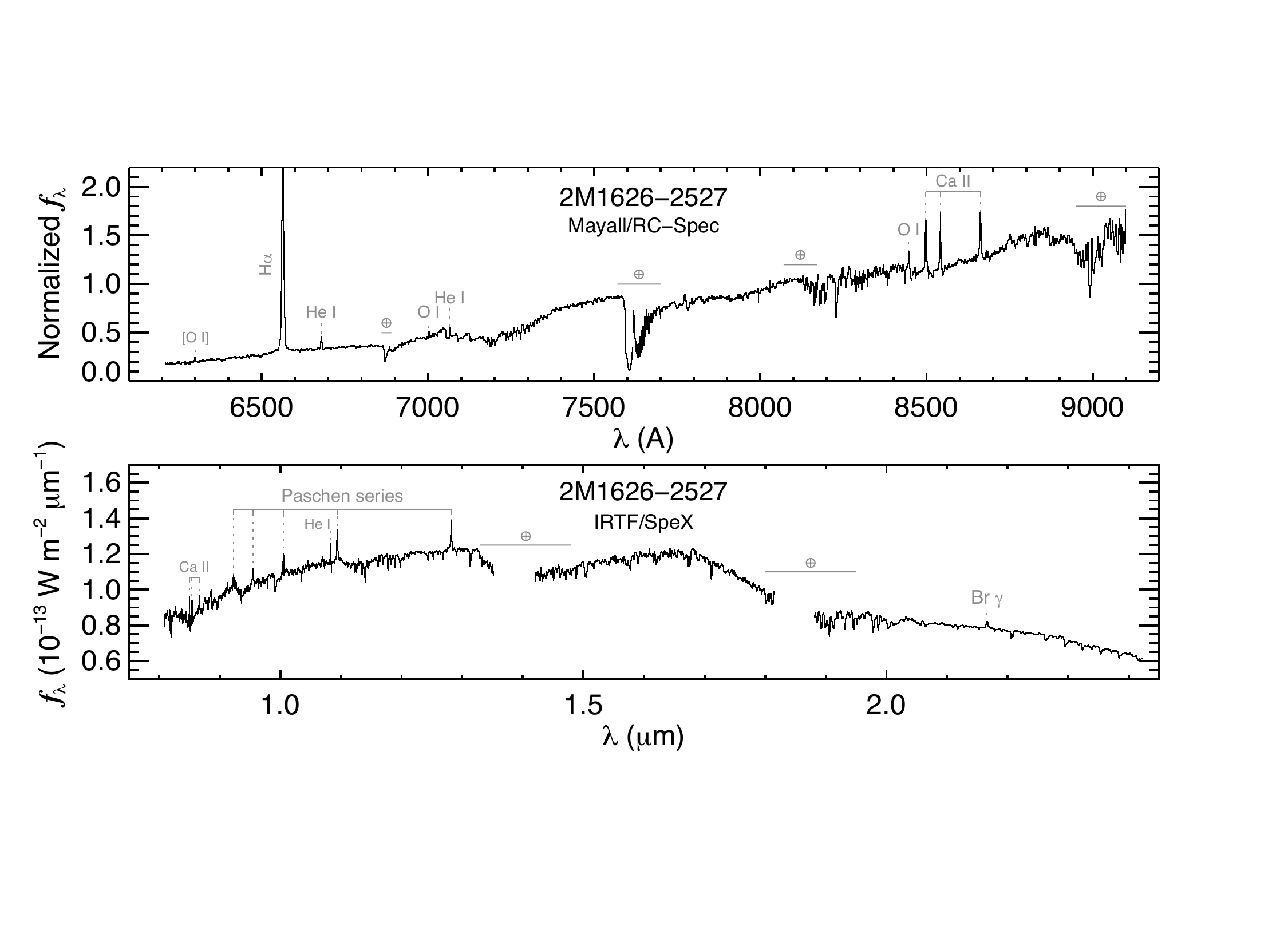}}
  \vskip -.9 in
  \caption{Optical and near-infrared spectra of 2M1626--2527 from Mayall/RC-Spec and IRTF/SpeX.  Emission
  lines from \ion{He}{1}, [\ion{O}{1}], \ion{Ca}{2}, \ion{He}{1}, and \ion{H}{1}--- including strong H$\alpha$ emission 
  (EW = --46~\AA)--- indicate active accretion onto this early-M dwarf.  Note that strong veiling is evident 
  at $\lambda$$\lesssim$7000~\AA.  \label{fig:optnirspecfig} } 
\end{figure*}

% Figure 14

\begin{figure}
  \vskip -.3 in
  \hskip -1.4 in
  \resizebox{6.2in}{!}{\includegraphics{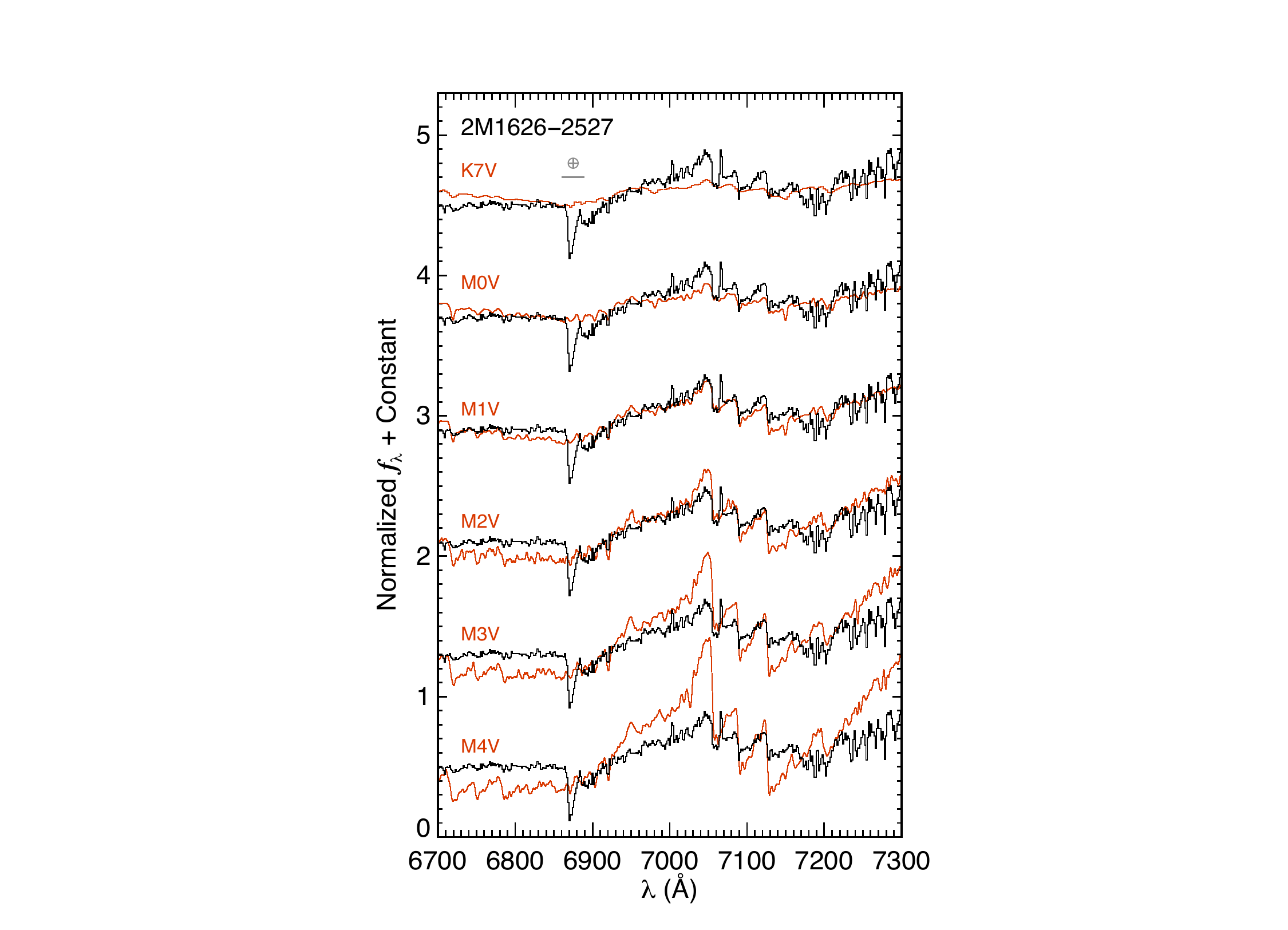}}
  \vskip -.1 in
  \caption{Comparison of our RC-Spec optical spectrum of 2M1626--2527 with K and M dwarf field templates
  from  \citet{Mann:2013fv} and \citet{Bochanski:2007it}.  The TiO bandheads between 7050--7150~\AA \ imply an
  early-M spectral type, with M1 being the best match.  Note that the absorption feature at $\approx$6870~\AA \ 
  is telluric (O$_2$ B band).  \label{fig:templatecomp} } 
\end{figure}

% Figure 15

\begin{figure*}
  \vskip -1.1 in
  \hskip 0. in
  \resizebox{7in}{!}{\includegraphics{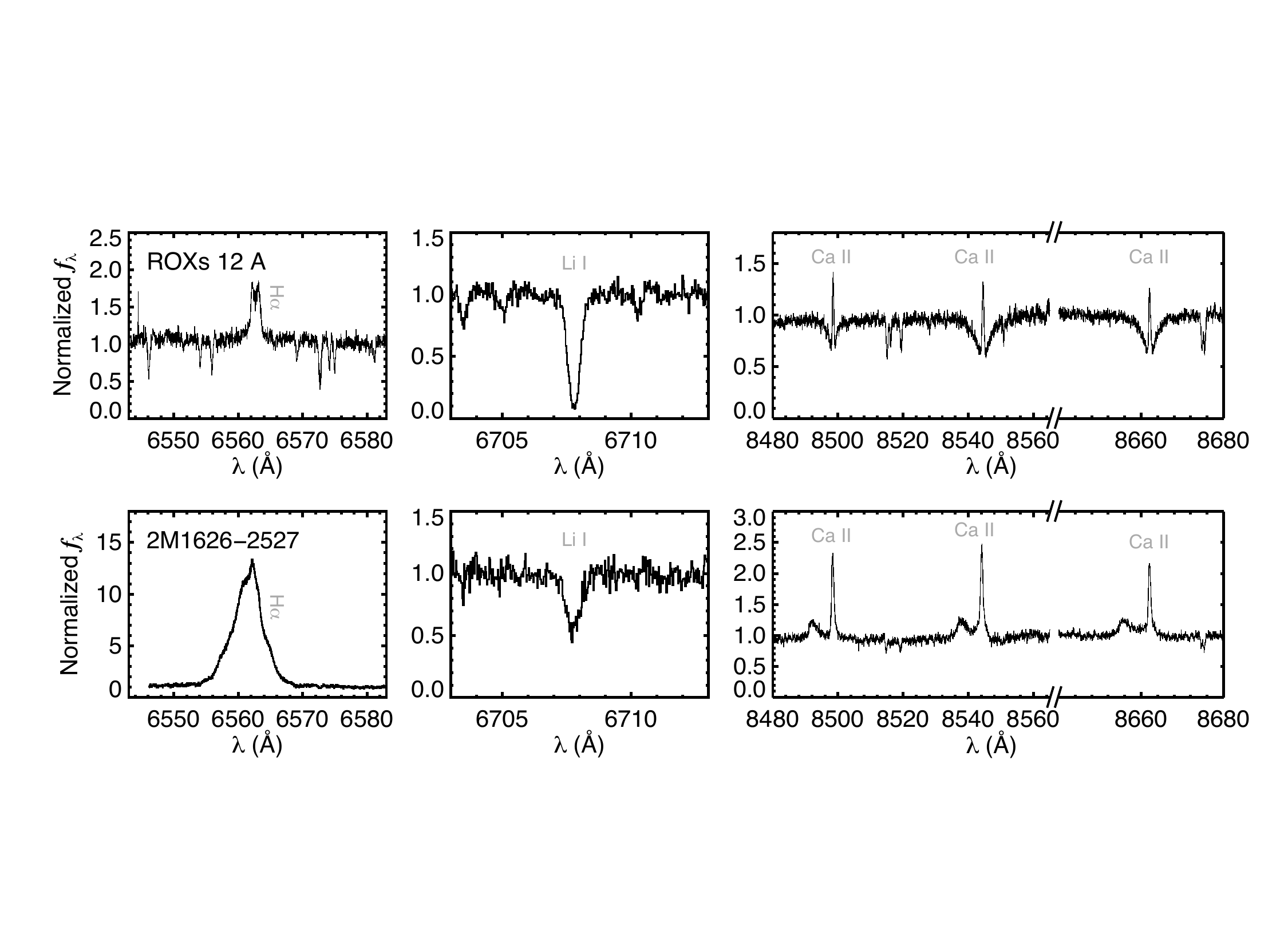}}
  \vskip -.9 in
  \caption{Keck/HIRES observations of ROXs 12 A (upper panels) and 2M1626--2527 (lower panels).  
  ROXs 12 A shows relatively weak H$\alpha$ emission (EW = --1.41 $\pm$ 0.02~\AA) despite strong lithium
  absorption (EW = 0.54 $\pm$ 0.01 \AA). 2M1626--2527 shows clear signs of high accretion rates, 
  with strong H$\alpha$ (EW = --53.17 $\pm$ 0.13 \AA) and modest, possibly veiled lithium (0.32 $\pm$ 0.02 \AA).  
  The calcium infrared triplet is in emission for both stars.   \label{fig:hires} } 
\end{figure*}

For ROXs 12 B, dereddened $J$, $H$, $Kp$, and $L'$ photometry from \citet{Kraus:2014tl} 
are in good agreement with the 2300 K BT-Settl model, which was chosen based on
spectral type-effective temperature relations for young brown dwarfs (see Section~\ref{sec:roxs12bphysical}).  
There is no evidence of excess emission from ROXs 12 B
that would indicate it hosts a circumsubstellar disk.  This contrasts with many other young brown dwarf 
and planetary-mass companions that harbor accretion 
subdisks as evidenced by optical/NIR emission lines and thermal infrared excess 
(e.g., \citealt{Bowler:2011gw}; \citealt{Bailey:2013gl}; \citealt{Zhou:2014ct}).

No spectral type has been published for 2M1626--2527.  Our own 
low-resolution optical spectrum shown in Figure~\ref{fig:optnirspecfig} reveals
a late-type spectrum with clear signs of active accretion including  
strong H$\alpha$ emission (EW=--46 \AA) as well as various transitions from \ion{He}{1}, [\ion{O}{1}], 
and \ion{Ca}{2}, all in emission (Table~\ref{tab:ew}).
Aside from  H$\alpha$ emission, the optical spectrum of 2M1626--2527 is relatively
featureless shortward of $\approx$7000~\AA \ and appears to be heavily veiled.
This likely originates from the
boundary layer of an accretion disk (e.g., \citealt{Basri:1990cc}; \citealt{Hartigan:1991fy}).
The near-infrared spectrum of 2M1626--2527 shown in Figure~\ref{fig:optnirspecfig} 
also points to a late (M-type) spectrum with strong emission lines from \ion{Ca}{2}, \ion{He}{1}, and \ion{H}{1} 
(the Paschen series and Br $\gamma$; see Table~\ref{tab:ew}).

The relative depths of the TiO bandheads spanning 7050--7150~\AA  \ 
are highly sensitive to temperature and can 
be used as an indicator of spectral type for 2M1626--2527 despite the heavy veiling at shorter wavelengths.
Figure \ref{fig:templatecomp} shows a detailed view of this region compared to 
K4--M4 templates from \citet{Bochanski:2007it} for the M dwarfs and \citet{Mann:2013fv} for the K dwarfs.
The relatively pronounced TiO bands in 2M1626--2527 are stronger than K4--M0 templates and weaker than M2--M4
templates, with M1 being the best match.  We therefore assign 2M1626--2527 a spectral type of M1
with an uncertainty of one subclass to reflect the imperfect agreement with this field template.
The corresponding effective temperature is 3720$^{+180}_{-160}$ K and 3630 $\pm$ 140 K using the 
\citet{Herczeg:2014is} and  \citet{Pecaut:2013ej} conversions, respectively.
We adopt $T_\mathrm{eff}$ = 3700 $\pm$ 150 K for 2M1626--2527.
A model effective temperature of 3700 K and reddening of $A_V$=1.8 mag is a good fit to
the photometry of 2M1626--2527 in the optical and near-infrared, but beyond about 3 $\mu$m
there is clearly very strong excess emission from a protoplanetary disk (Figure \ref{fig:sedfig}).

A few selected regions of our HIRES spectra of ROXs 12 A and 2M1626--2527 are shown in Figure~\ref{fig:hires}.
ROXs 12 A shows relatively weak H$\alpha$ emission ($EW$=--1.41 $\pm$ 0.02 \AA) 
considering its young age.  This is similar to the value of 1.2~\AA \ found by \citet{Bouvier:1992ww} and
--2.00 $\pm$ 0.02 \AA \ from \citet{Rizzuto:2015bs}.
These values are consistent with the lower envelope of H$\alpha$ emission for pre-main sequence M0 stars 
spanning the youngest star-forming regions like Taurus (\citealt{Kraus:2017bg}) to 
somewhat older populations like Upper Scorpius (\citealt{Rizzuto:2015bs}).  
2M1626--2527 shows strong, broadened H$\alpha$ emission ($EW$=--53.17 $\pm$ 0.13 \AA) in our HIRES spectrum, 
comparable to our measurement with RC-Spec ($EW$=--46.5 $\pm$ 0.5 \AA).
Both stars show \ion{Li}{1} $\lambda$6708~\AA \ absorption; we measure an equivalent width of 
0.54 $\pm$ 0.01 \AA \ for ROXs 12 A--- in good agreement with the value of 0.52 $\pm$ 0.02 \AA \ from
\citet{Rizzuto:2015bs}--- and $EW$(Li) = 0.32 $\pm$ 0.02 \AA \ for 2M1626--2527.
The \ion{Ca}{2} infrared triplet is also seen in emission for both sources, with ROXs 12 A exhibiting
broad absorption with superimposed central cores in emission.  The \ion{Ca}{2} line profiles from
2M1626--2527 show a broad, blueshifted component in emission, 
offset from the central narrow line emission.  
This suggests that we are viewing a face-on accretion disk producing a broadened component along our line of sight
together with a narrow zero-velocity peak where material hits the star.  This geometry is consistent with
the mostly pole-on stellar inclination we find for 2M1626--2527 in Section~\ref{sec:inclinations}.

\subsection{$K2$ Light Curves}{\label{sec:k2periods}}

% Figure 16

\begin{figure*}
  \vskip -.8 in
  \hskip 0 in
  \resizebox{7.3in}{!}{\includegraphics{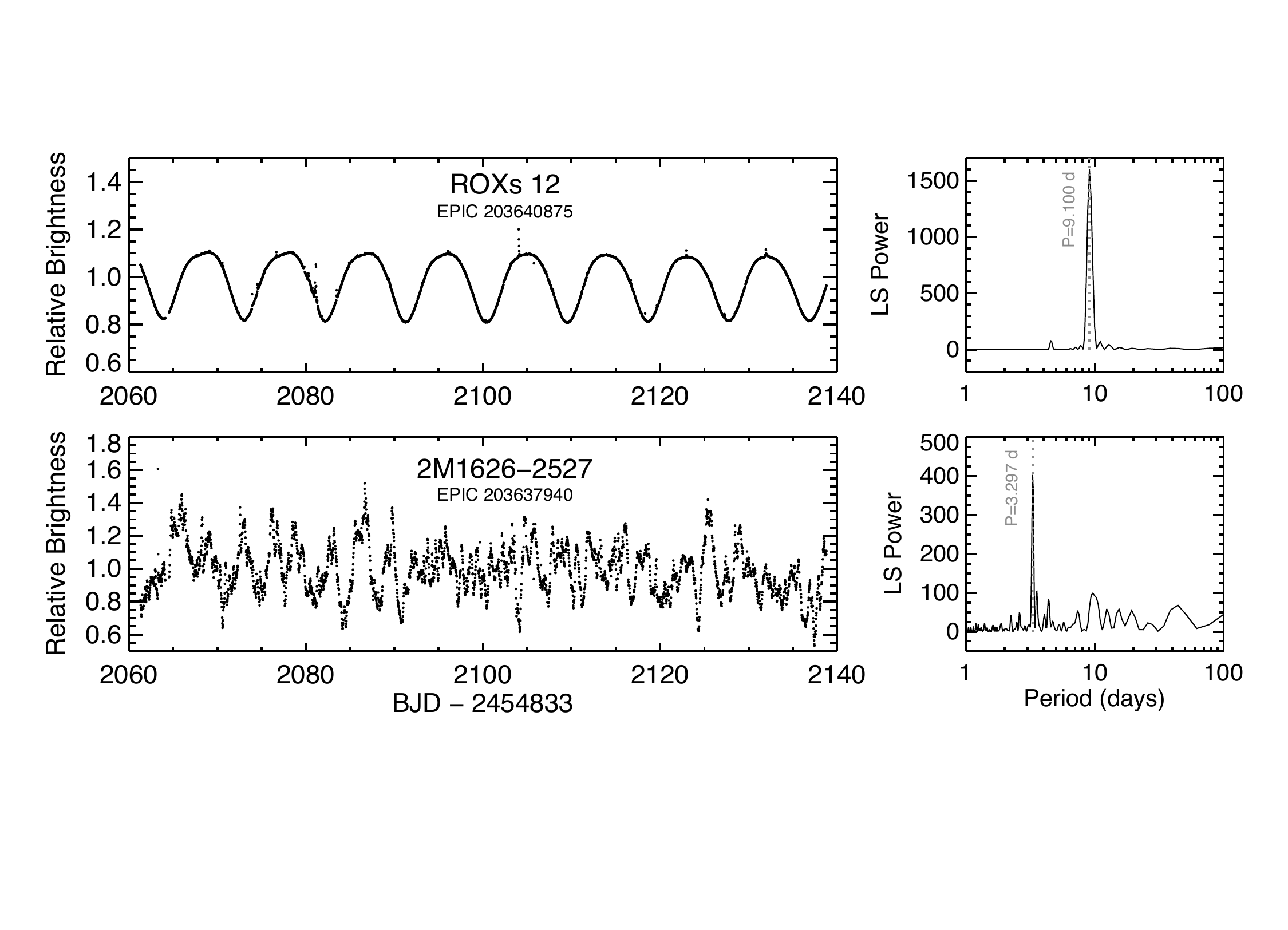}}
  \vskip -1.2 in
  \caption{$K2$ light curves of ROXs 12 A and 2M1626--2527.  ROXs 12 A displays regular, large-amplitude modulations 
  with a period of 9.1 days, as seen in the Lomb-Scargle periodogram in the right panels.
  2M1626--2527 shows stochastic variability pointing to active accretion.  The periodogram for
  this star shows a significant peak at 3.297 days which we interpret as underlying rotational modulation.
  Phased light curves are plotted in Figure~\ref{fig:k2phased}.    \label{fig:k2} } 
\end{figure*}

% Figure 17

\begin{figure}
  \vskip -.35 in
  \hskip -0.6 in
  \resizebox{4.2in}{!}{\includegraphics{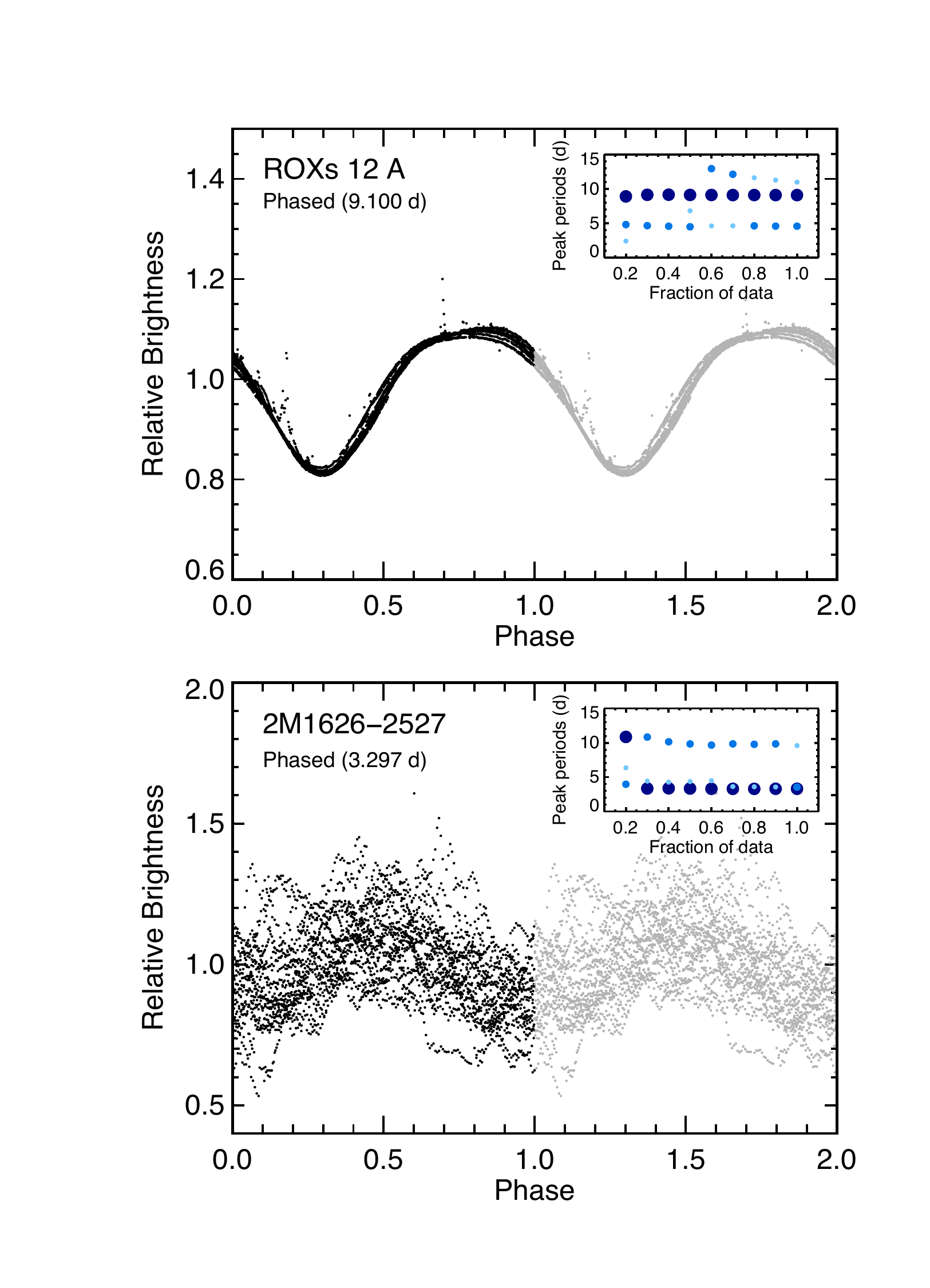}}
  \vskip -.25 in
  \caption{$K2$ light curves of ROXs 12 A and 2M1626--2527 phased to the highest periodogram peak.  ROXs 12 A
  shows slight changes from period to period, likely caused by slowly-evolving starspots.  Several flares are also evident.
  2M1626--2527 is much more variable, but an underlying periodic signature is clearly visible in this phased curve.
  The inset panels for each star show the period of the highest three periodogram peaks (large, medium, and small 
  filled circles, respectively) as a function of the fraction of data being considered.  For ROXs 12 A, a consistent $\approx$9 d period is 
  the highest periodogram signal spanning the first 20\% of the data out to the full light curve.
  For 2M1626--2527, the peak period for first 20\% of the data is at $\approx$11 d, but drops down to
  $\approx$3.3 d thereafter.  This suggests that the underlying 3.3 d signal is probably not spurious as it remains
  consistent over time.    \label{fig:k2phased} } 
\end{figure}

The $K2$ light curves for ROXs 12 A and 2M1626--2527 are shown in Figure \ref{fig:k2}.
ROXs 12 A exhibits remarkably strong periodic changes with peak-to-peak semi-amplitude 
variations of $\approx$14\%.  The most straightforward interpretation of these modulations
is that they are caused by large, long-lived regions of inhomogeneous spot coverage coming in and out of view
as the star rotates.  A Lomb-Scargle (LS) periodogram
(\citealt{Lomb:1976wz}; \citealt{Scargle:1982eu}; \citealt{Horne:1986ds}) 
reveals a strong peak period of 9.0998 days, which
we adopt as the rotation period of ROXs 12 A.  This is near the upper envelope of (but consistent with) 
rotation period measurements of low-mass pre-main sequence stars during the 
first $\approx$1--10~Myr of their evolution 
(e.g., \citealt{Rebull:2004ed}; \citealt{Irwin:2011gz}; \citealt{Bouvier:2014ik}).
Several flaring events are clearly visible over the course of these observations. 
The $K2$ light curve of ROXs 12 A phased to 9.100 d (Figure~\ref{fig:k2phased})
exhibits only slight changes in the shape of the light curve from period to period, indicating
minimal evolution of starspots on timescales less than one period.

The light curve of 2M1626--2527 shows both low-amplitude (few percent) and 
high-amplitude ($\approx$40\%) variations on timescales ranging from hours to days (Figure \ref{fig:k2}).
This stochastic behavior is characteristic of young accreting stars and is thought to be caused by 
time-dependent mass accretion events that produce transient hot spots on the stellar photosphere
(e.g., \citealt{Cody:2014el}; \citealt{Stauffer:2016cu}).
\citet{Cody:2017ct} find this phenomenon is fairly common, occurring among $\approx$9\% of 
$\rho$ Oph and Upper Sco members with strong infrared excesses.
Although there are no obvious signs of periodicity in the light curve, 
the LS periodogram shows a significant peak power at 3.297 days.  
To explore whether this represents a real underlying modulation superimposed on stochastic variations,
we applied the same periodogram analysis to smaller portions of the light curve which should
return the same approximate peak period near 3.3 days
if the modulations are authentic and consistent over time.  
The inset plot in the lower panel of Figure~\ref{fig:k2phased} shows the peak period, second-highest peak,
and third-highest peaks starting with 20\% of the light curve and progressively including more data
in 10\% bins.  Beginning with 30\% of the data, the peak periods range between 3.28 and 3.37 d.
This suggests the underlying modulations are real and are probably caused by rotationally modulated starspots.
Interestingly, this means that 2M1626--2527 is rotating about three times faster than ROXs 12 A
despite having an accreting disk, which has been shown on average to slow the rotation rates of young
stars via star-disk interactions (e.g., \citealt{Herbst:2002je}; \citealt{Cieza:2007bi}).
This is unusual but not entirely unexpected given the broad spread in  periods for diskless and
disk-bearing young stars.

We estimate uncertainties for our light curve periods following \citet{Kovacs:1981bs} and \citet{Horne:1986ds}.
For a single signal with Gaussian noise, they found that the error in the frequency measurement $\nu$ is 
$\delta$$\nu$ = 3$\sigma$/(4$\sqrt N$$T$$A$), where $\sigma$ is the standard deviation of the noise about
the signal, $N$ is the number of data points, $T$ is the period, and $A$ is the signal semi-amplitude.  From this the 
period ($P$) uncertainty is $\delta$$P$ = $P^2$$\delta$$\nu$.
We fit a sine curve to the phased data to find $A$ from the $K2$ light curves.  
The rms noise was approximated by computing the standard deviation of the residuals between the time series
photometry and a Gaussian smoothed version of the same data.
From this analysis we find formal uncertainties of 0.009 d for ROXS 12 and 0.002 d for 2M1626--2527.
The high precision of these errors is primarily a result of the large number of periods sampled during the
observing window (9--24 full cycles), the large number of data points in these light curves ($\approx$3400), and 
the relatively high semi-amplitudes for both systems (10--14\%).

Stars exhibit differential rotation between their poles and equator which can make it difficult to
infer the equatorial rotation period from light curves alone if the location of their starspots is unknown.  
For example, the Sun has a longer rotation period of about 34 d at the poles compared to 25 d at its equator.  
Our period measurements for ROXs 12 A and 2M1626--2527 may therefore be over-estimating
the \emph{equatorial} rotation periods of these objects depending on which latitudinal regions the starspots were located.
\citet{Reinhold:2015ep} show that the  pole-equatorial shear 
($\Delta$$\Omega$ = 2$\pi$(1/$P_\mathrm{min}$ -- 1/$P_\mathrm{max}$) )
of $Kepler$ stars with comparable effective temperatures to ROXs 12 A and 2M1626--2527 ranges from
about 0.01 to 0.10 rad/d, with a typical value of about 0.03 rad/d.  
This typical shear would imply a range of maximum and minimum periods of 
8.7--9.5 d for ROXs 12 A and 3.25--3.35 d for 2M1626--2527.
To take into account possible effects of differential rotation,
we therefore adopt larger rotation period uncertainties
of 9.1 $\pm$ 0.4 d for ROXs 12 and 3.30 $\pm$ 0.05 d for 2M1626--2527
when using these values in the context of \emph{equatorial} rotation period estimates.

\subsection{Stellar Luminosities, Masses, and Ages}{\label{sec:lum}}

Bolometric luminosities for ROXs 12 A and 2M1626--2527 are derived using the
3900 K and 3700 K BT-Settl models, respectively, shown in Figure~\ref{fig:sedfig}.  
The models are first scaled to the $H$-band de-reddened flux and then integrated 
from 0.2--200 $\mu$m to find the bolometric flux.  
A distance of 137 pc is assumed based on VLBA parallax measurements
of Ophiuchus members in the Lynds 1688 dark cloud (\citealt{OrtizLeon:2017ce}),
and we adopt a $\pm$ 10 pc uncertainty to encompass ambiguity in Ophiuchus versus Upper Sco membership.
Uncertainties are derived by integrating models 100 K warmer and 100 K cooler
than the nominal stellar effective temperatures and calculating the mean difference
between these and the stars' bolometric fluxes.
Surface gravities of $\log g$ = 4.0 dex are chosen based on expectations from evolutionary models
(e.g., \citealt{Baraffe:2015fwa}), as substantial deviations from this are unphysical.
This yields luminosities of $\log L_\mathrm{bol}/L_{\odot}$ = --0.57 $\pm$ 0.06 dex for ROXs 12 A
and --0.81 $\pm$ 0.06 dex for 2M1626--2527.

Stellar masses and ages are estimated using \citet{Baraffe:2015fwa} evolutionary models.
For a given effective temperature and luminosity, we identify the corresponding mass and age
by finely interpolating the grid of evolutionary models (Figure~\ref{fig:radlum}).  This process is repeated in a Monte Carlo
fashion to build a distribution of masses and ages assuming normally-distributed errors in 
$T_\mathrm{eff}$ and log($L_\mathrm{bol}$/$L_{\odot}$).
We infer a mass of 0.65$^{+0.05}_{-0.09}$ \Msun \ and an age of 6$^{+4}_{-2}$ Myr for ROXs 12 A.
For 2M1626--2527 we find a mass of 0.50$^{+0.10}_{-0.10}$ \Msun \ and an age of 8$^{+7}_{-4}$ Myr.

\section{Discussion}{\label{sec:discussion}}

\subsection{Ophiuchus or Upper Scorpius?}{\label{sec:ophusco}}

The question of whether this system belongs to the younger Ophiuchus association or the older, more expansive
Upper Sco region has  implications for the inferred mass of 
the substellar companion ROXs 12 B.  
ROXs 12 has historically been regarded as a member of 
the $\approx$0.5--2~Myr Ophiuchus star-forming region since its identification by \citet{Bouvier:1992ww}.
This region is centered on the dark cloud Lynds 1688 and its $>$300 members comprise 
a range of evolutionary stages spanning embedded protostars,
accreting T Tauri stars, transition disks, and an older surface population that merges with the 
$\approx$5--10 Myr Upper Scorpius subgroup (\citealt{Wilking:2008wq}).
ROXs 12 and 2M1626--2527 are located about 0.4$^{\circ}$ from the L1688 cloud core
in (or behind) an extended region of modest extinction ($A_V$ = 1--2 mag; \citealt{Cambresy:1999vv}).

Unfortunately, 
because the internal velocity dispersions of cluster members are
$\approx$1.0 km s$^{-1}$ ($\approx$1.5 mas yr$^{-1}$; \citealt{Kraus:2008ix}; \citealt{Wilking:2015hn})
the kinematics of Oph and Upper Sco are nearly indistinguishable 
using current proper-motion surveys.
\citet{Mamajek:2008dx} finds a mean proper motion of $\mu_{\alpha}$cos$\delta $ = --10 $\pm$ 1.5 mas yr$^{-1}$
and $\mu_{\delta}$ = --27 $\pm$ 1.5 mas yr$^{-1}$ (1 mas yr$^{-1}$ systematic error) for 
Oph cloud members.
Based on the $UVW$ space velocities of Upper Sco from \citet{Rizzuto:2011gs},
the expected proper motion of an Upper Sco member at the sky position of ROXs 12 is 
$\mu_{\alpha}$cos$\delta $ = --11.9 $\pm$ 1.5 mas yr$^{-1}$ and $\mu_{\delta}$ = --24.4 $\pm$ 1.5 mas yr$^{-1}$.
(Note that because of the large angular extent of Upper Sco on the sky, the proper motions of its members
vary slightly across this region so we use the association's three-dimensional space velocity at the sky position of ROXs 12 
to calculate the expected proper motion.)
The measured proper motions of ROXs 12 A and 2M1626--2527 (Table \ref{tab:properties}) 
agree with both of these regions to within about 2 $\sigma$.
Unsurprisingly, we find Sco-Oph membership probabilities of 99\% for both stars 
following the Bayesian membership analysis described in \citet{Rizzuto:2011gs} and \citet{Rizzuto:2015bs}.

Regardless of kinematics, the ages inferred for ROXs 12 A and 2M1626--2527 from their 
positions on the HR diagram are most consistent with Upper Sco.  
\citet{Pecaut:2016fua} adopt a median age of 10 $\pm$ 3 Myr for Upper Sco with a large intrinsic age spread of
$\pm$7 Myr, which is consistent with recent findings by \citet{Fang:2017ix}.  Pecaut et al. also find evidence of an age gradient in this subgroup, with ROXs 12 positioned
near the transition point between $\approx$5--9 Myr ages in the northwest and $\approx$11--15 Myr ages in the south-east.
Our independently-inferred ages for ROXs 12 A (6$^{+4}_{-2}$ Myr) and 2M1626--2527 (8$^{+7}_{-4}$ Myr) from
Section~\ref{sec:lum} (also see Figure~\ref{fig:radlum}) appear to 
agree with typical ages of Upper Sco members in the vicinity of the Ophiuchus cloud.
In the near future, this membership analysis and age will be refined with parallactic distances and precise proper motions
from \emph{Gaia}.

% Figure 18

\begin{figure}
  \vskip -2.4 in
  \hskip -.6 in
  \resizebox{4.2in}{!}{\includegraphics{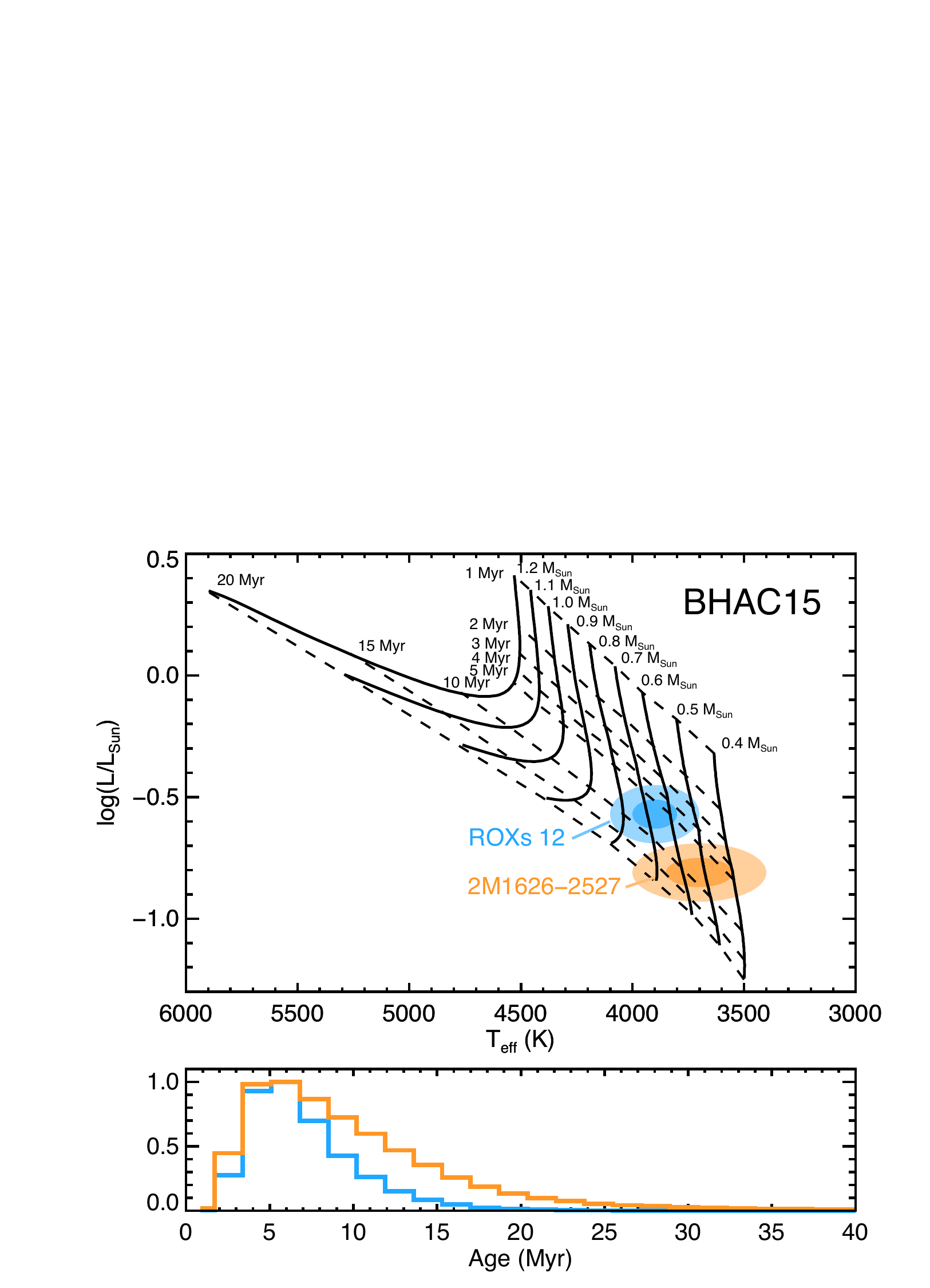}}
  \vskip -.1 in
  \caption{Comparison of ROXs 12 A (blue) and 2M1626--2527 (orange) to \citet{Baraffe:2015fwa} evolutionary models.
    \emph{Upper panel:} H-R diagram showing the location of both stars from their effective temperatures and bolometric 
    luminosities.  
    Dark shaded ellipses show the 1 $\sigma$ confidence region; light shaded ellipses 
    show the 2 $\sigma$ region.  The corresponding masses from these models are 0.65$^{+0.05}_{-0.09}$ \Msun \
    and 0.5$^{+0.1}_{-0.1}$ \Msun \ for  ROXs 12 A and 2M1626--2527, respectively.
    \emph{Lower panel:} Age distributions for both stars from their positions on the H-R diagram.  The inferred age for 
    ROXs 12 A is 6$^{+4}_{-2}$ Myr and for 2M1626--2527 is 8$^{+7}_{-4}$ Myr.  Distributions 
    have been normalized to their peak values for visual purposes.       \label{fig:radlum} } 
\end{figure}

\subsection{Stellar Radii and Line-of-Sight Inclinations}{\label{sec:inclinations}}

The radii of ROXs 12 A and 2M1626--2527 can be determined with the 
Stefan-Boltzmann law using their effective temperatures and bolometric luminosities.
This yields values of 1.14 $\pm$ 0.09 $R_{\odot}$ for ROXs 12 A and 
0.96 $\pm$ 0.10 $R_{\odot}$ for 2M1626--2527, where uncertainties are
propagated analytically.

A lower limit on the radius of ROXs 12 A can also be inferred from the measured 
rotation period ($P_\mathrm{rot}$) and projected rotational velocity 
($v_p$ = $v$sin$i_*$).  
By equating $v$ and 2$\pi$$R_*$/$P_\mathrm{rot}$, 
the stellar radius $R_*$ and associated uncertainty $\sigma_{R_*}$ are

\begin{equation}{\label{eq:rad}}
R_* (i_*) = \frac{P_\mathrm{rot} v_p}{2 \pi \sin i_*}
\end{equation}

\begin{equation}
\sigma_{R_*} \approx R_* \bigg( \Big( \frac{\sigma_{P_\mathrm{rot}} }{P_\mathrm{rot}} \Big)^{2} +  \Big( \frac{\sigma_{v_p} }{v_p} \Big)^{2} + 
 \Big( \sigma_{i_*} \frac{\cos i_*}{\sin i_*} \Big)^{2}  \bigg)^{1/2}.
\end{equation}

% Figure 19

\begin{figure}
  \vskip -.2 in
  \hskip -.6 in
  \resizebox{4.5in}{!}{\includegraphics{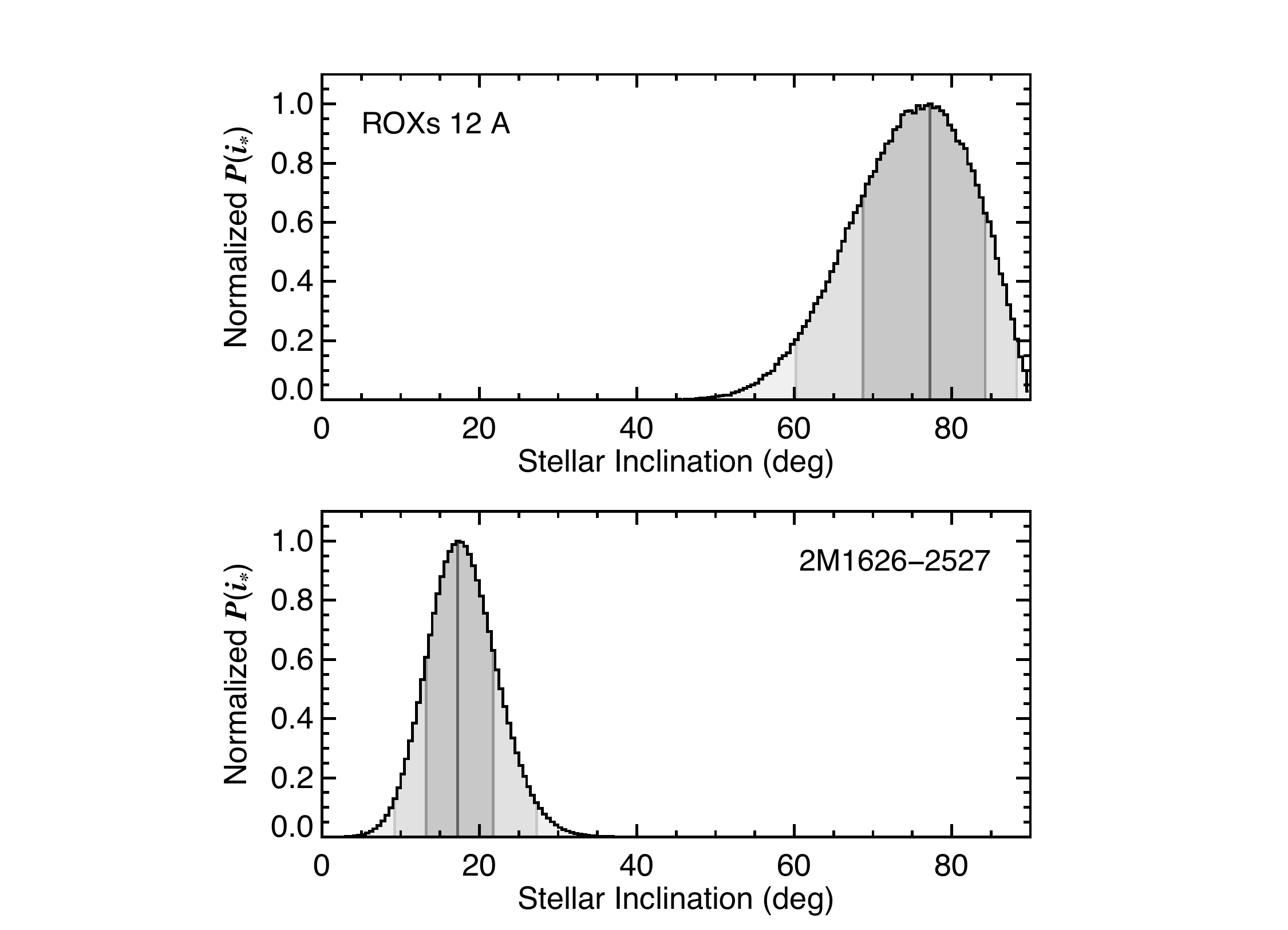}}
  \vskip -.1 in
  \caption{Constraints on the line-of-sight inclination of ROXs 12 A and 2M1626--2527
  based on the stars' rotation periods, projected rotational velocities, effective temperatures, and bolometric
  luminosities (Equation~\ref{eq:incl}).  ROXs 12 A is viewed mostly edge-on, 
  with an inclination distribution that peaks at 77$^{\circ}$ and a 95.4\% highest posterior density 
  interval of 60--88$^{\circ}$.  2M1626--2527 is viewed mostly pole-on, 
  with an inclination distribution that peaks at 17$^{\circ}$ and a 95.4\% highest posterior density 
  interval of 9--27$^{\circ}$.   \label{fig:inclfig} } 
\end{figure}

In cases where the inclination is unknown, Equation~\ref{eq:rad} becomes a lower limit 
on $R_*$ if an edge-on orbit is assumed ($i_*$=90$^{\circ}$).
Adopting $P_\mathrm{rot}$ = 9.1 $\pm$ 0.4 d and $v_p$ = 8.2 $\pm$ 0.7 km s$^{-1}$ for ROXs 12 A
implies $R_*$ $>$ 1.47 $\pm$ 0.14 $R_{\odot}$.
There is tension between this radius and the radius implied from the Stefan-Boltzmann law at the
2.0 $\sigma$ level.  The most likely origin of this discrepancy is (1) that the inferred effective temperature for ROXs 12 A is
too high,  (2) our $v$sin$i_*$ measurement is too high, 
and/or (3) the rotation period samples high-latitude rather than equatorial starspots.

The line-of-sight inclination of ROXs 12 A can be determined more precisely 
using \emph{joint} constraints from the ``spectroscopic'' (Stefan-Boltzmann) radius and the
``rotational'' ($R_*$($i_*$)) radius.  Equating these two radii and solving for the stellar inclination gives

\begin{equation}{\label{eq:incl}}
i_* = \sin^{-1} \Big( \sqrt \frac{\sigma_\mathrm{SB}}{\pi  L_\mathrm{bol}}  v_p P_\mathrm{rot} T_\mathrm{eff}^2  \Big),
\end{equation}

\noindent where $\sigma_\mathrm{SB}$ is the Stefan-Boltzmann constant.
Uncertainties are incorporated in an Monte Carlo fashion to build a line-of-sight inclination distribution,
which peaks at 77$^{\circ}$ and is truncated at 90$^{\circ}$ (Figure \ref{fig:inclfig}).  
Monte Carlo realizations that result in $\sin i_*$ values $>$1 are excluded.
The mode and 68.3\% (1-sigma equivalent) highest posterior density interval is 77$^{+7}_{-9}$$^{\circ}$.
The 95.4\% (2-sigma equivalent) range spans 60--88$^{\circ}$.  Note that these line-of-sight stellar inclinations are 
symmetric about 90$^{\circ}$ and produce mirrored observational signatures at higher inclinations.

Similarly, using $P_\mathrm{rot}$ = 3.30 $\pm$ 0.05 d and $v_p$ = 4.5 $\pm$ 1.0 km s$^{-1}$ for 2M1626--2527
implies $R_*$ $>$ 0.29 $\pm$ 0.07 $R_{\odot}$.  This agrees with the spectroscopic radius of 0.96 $\pm$ 0.10 $R_{\odot}$.
The joint constraint on the inclination distribution for 2M1626--2527 is shown in Figure \ref{fig:inclfig}.
The mode and 68.3\% highest posterior density interval is 17$^{+5}_{-4}$$^{\circ}$, and 
the 95.4\% range spans 9--27$^{\circ}$.

The difference between these two inclination distributions, $\Delta i_*$, is the degree to which these
two stars are misaligned.  
For ROXs 12 A and 2M1626--2527 we find $\Delta i_*$ = 60$^{+7}_{-11}$$^{\circ}$, indicating these stars
are strongly misaligned.  This is not  surprising for two young stars separated by over 5000~AU, and it adds to 
mounting evidence that misalignments between the components of wide binaries is a common phenomenon 
(e.g., \citealt{Hale:1994gv}; \citealt{Jensen:2014hv}; \citealt{Williams:2014fg}).
This indicates that either the initial fragmenting cloud core that produced this system did not
act as a uniformly co-rotating collapsing body, that this wide companion was captured, or that 
the rotational inclinations of these systems to cause them to evolve with time.

\subsection{Obliquity of ROXs 12 AB}{\label{sec:obliquity}}

% Figure 20

\begin{figure}
  \vskip -.4 in
  \hskip -.7 in
  \resizebox{4.4in}{!}{\includegraphics{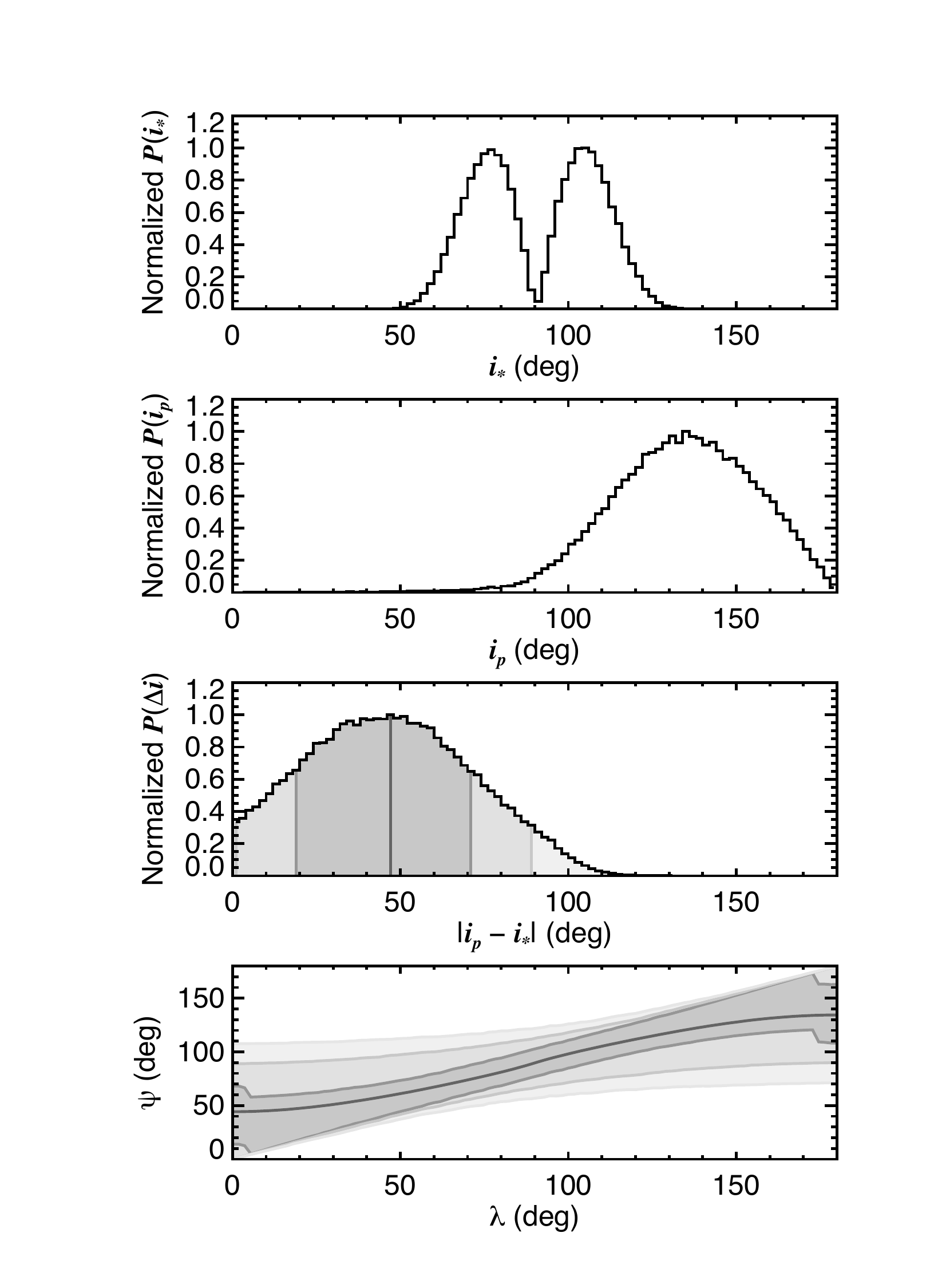}}
  \vskip -.25 in
  \caption{\emph{Top:} Line-of-sight inclination ($i_*$) of ROXs 12 A.  
  Here we have mirrored the distribution from Figure~\ref{fig:inclfig} about 90$^{\circ}$ owing to
  ambiguity in viewing geometry, which produces bimodal peaks.  
  \emph{Upper Middle:} Orbital inclination ($i_p$) of ROXs 12 B
  from \citet{Bryan:2016eo} based on measured orbital motion of the companion.  
  \emph{Lower Middle:} Absolute difference between the stellar rotational axis and 
  orbital axis of the companion ($\Delta i$ = $|i_p - i_*|$).  This difference is a lower bound on the true deprojecrted 
  obliquity angle ($\psi$).  The distribution peaks at 49$^{\circ}$ with a 1-$\sigma$ 
  minimum confidence interval of 17--69$^{\circ}$, indicating ROXs 12 A and B likely
  exhibit spin-orbit misalignment. 
  \emph{Bottom:} Contour plot demonstrating how the obliquity angle constraint 
  varies as a function of the sky-projected spin-orbit angle ($\lambda$) for the ROXs 12 AB system.  
  The slice at $\lambda$=0$^{\circ}$ corresponds to the distribution presented in the lower middle panel of this figure.
  At greater values of (the unknown angle) $\lambda$, the distribution of $\psi$ values steadily increases to higher values.
  Contours represent 1, 2, and 3-$\sigma$ minimum confidence intervals.
     \label{fig:obliquity} } 
\end{figure}

ROXs 12 B has a projected  separation of 240 AU, which corresponds to an orbital period of 
4600 yr and angular orbital motion of 0.08$^{\circ}$ yr$^{-1}$ assuming a circular face-on orbit.
Because of the long  baseline since the initial discovery epoch of ROXs 12 B by 
\citet{Ratzka:2005jv} in 2001, enough time elapsed for \citet{Kraus:2014tl} and \citet{Bryan:2016eo} to 
detect small but significant orbital motion from this companion.  
Despite the small amount of orbital coverage, \citet{Bryan:2016eo} were able to constrain
the orbital elements of ROXs 12 B using the efficient rejection sampling algorithm for Keplerian
orbits described in \citet{Blunt:2017eta}.

Of particular interest for this study is the posterior distribution of the orbital inclination, $i_p$,
which peaks at about 135$^{\circ}$ with a broad tail extending from about 90$^{\circ}$ to 180$^{\circ}$,
as shown in Figure~\ref{fig:obliquity}.   
The inclination of the stellar rotation axis of ROXs 12 A, $i_*$, can be compared with $i_p$
to offer clues about the stellar obliquity--- the relative orientation of the stellar spin axis and 
orbital angular momentum vector.
The \emph{true} (de-projected) angle between the stellar spin and planetary orbital axes, $\psi$, 
is related to the sky-projected spin-orbit angle, $\lambda$, as follows:

\begin{equation}{\label{eq:psi}}
\psi = \cos^{-1} \Big( \cos{i_*} \cos{i_p} + \sin{i_*} \sin{i_p} \cos{\lambda} \Big).
\end{equation}

\noindent Diagrams showing the geometric configuration can be found in, e.g., \citet{Ohta:2005ux} and \citet{Fabrycky:2009ke}.
The three angles $\lambda$, $i_*$, and $i_p$ must be known to determine $\psi$.
The projected obliquity ($\lambda$) is regularly measured for large transiting planets 
using the Rossiter-McLaughlin effect (e.g., \citealt{Queloz:2000ui}; \citealt{Winn:2005tg}; \citealt{Johnson:2009jr}).
In those cases, $i_p$$\approx$90$^{\circ}$, so 
$\cos \psi \approx \sin{i_*} \cos{\lambda} $ and therefore the projected obliquity ($\lambda$) 
is a measurement of lower bound on the true obliquity ($\psi$) assuming $i_*$ is unknown.
Similarly, if $\lambda$ is unknown as is the case for ROXs 12 AB, the lower limit on $\psi$ can be determined from
the absolute difference between $i_p$ and $i_*$: $\psi$ $\ge$ $|{i_* - i_p}|$ $\equiv$ $\Delta i$ (see Appendix).
Therefore, a system can have spin-orbit misalignment if $\Delta i$ is zero, but a non-zero value
of $\Delta i$ means the system must be misaligned by at least that amount (\citealt{Glebocki:1997ul}).
In this sense, knowing both $i_p$ and $i_*$ offers comparable information about
$\psi$ as do measurements of the Rossiter-McLaughlin effect.  

The probability distribution functions for $i_*$, $i_p$, and $\Delta i$ for ROXs 12 A and B are shown in Figure \ref{fig:obliquity}.
Note that since $i_*$ is symmetric about 90$^{\circ}$ it has been mirrored from that of Figure~\ref{fig:inclfig} and
is therefore bimodal.
The distribution of $\Delta i$ values peaks at 49$^{\circ}$ with a 1-$\sigma$ minimum confidence interval of 
17--69$^{\circ}$.  From this, we calculate that the probability that $\Delta i$ is greater than 10$^{\circ}$ is 94\%.
This distribution represents the \emph{minimum} values of true obliquity angles in this system and indicates that
ROXs 12 A and B are likely to have spin-orbit misalignment, although alignment ($\Delta i$=0$^{\circ}$) 
cannot be ruled out from the observations.

\subsection{How Common is Pa$\beta$ Emission?}{\label{sec:pabeta}}

The presence of disks around
widely-separated brown dwarf and planetary-mass companions offers a convenient way to directly
study the formation and late stages of growth of these objects.  
These subdisks produce a variety of observational signatures: ultraviolet excess
emission from accretion of hot gas (\citealt{Zhou:2014ct}); hydrogen line emission 
like H$\alpha$ and Pa$\beta$ (e.g., \citealt{Bowler:2011gw}; \citealt{Wu:2015kh}); and
thermal excess emission in the mid-IR (\citealt{Bailey:2013gl}).
 Sub-mm emission from warm dust has only been detected
 for the ambiguous companion to FW~Tau (\citealt{Kraus:2015fx}; \citealt{Caceres:2015hg}) despite 
 ongoing searches targeting several other disk-bearing substellar companions 
 (e.g., \citealt{Bowler:2015hx}; \citealt{MacGregor:2017fl}; \citealt{Wu:2017kd}). 

Although we find no convincing evidence that ROXs 12 B harbors a subdisk,
the lack of Pa$\beta$ emission in this individual object can nevertheless be used to more broadly 
assess the global frequency of low-mass companions
with accretion rates large enough to result in Pa$\beta$ line emission.
The sample of young companions near and below the deuterium burning limit that also have
moderate-resolution $J$-band spectroscopy has  increased over the past few years 
to the point where we can begin to quantify the statistical properties of these objects
as a population.
Note that the strength of the Pa$\beta$ emission line is expected to vary strongly with effective temperature, which alters the
pseudocontinnuum level, as well as mass accretion rate.  In addition, both the S/N and resolving power of a spectrum can
influence the detection of an emission line.
For this analysis we ignore these effects and simply count reported Pa$\beta$ detections and nondetections
as discrete Bernoulli trials.

If we isolate a sample of young companions with ages $\lesssim$15~Myr, when such subdisks might
still be expected to be actively accreting, and companion masses $\lesssim$20~\Mjup,
there are 11 low-mass brown dwarfs and planetary-mass objects with published 
moderate-resolution spectroscopy.
Five of these show Pa$\beta$ in emission: 
GSC 6214-210 B (\citealt{Bowler:2011gw}; \citealt{Lachapelle:2015cx}),
CT Cha B (\citealt{Schmidt:2008bf}; \citealt{Bonnefoy:2014dh}),
GQ Lup B (\citealt{Seifahrt:2007iq}),
DH Tau B (\citealt{Bonnefoy:2014dh}; \citealt{Wolff:2017ky}),
and FW Tau b (\citealt{Bowler:2014dk}).
Six companions do not appear to have Pa$\beta$ in emission:
1RXS J1609--2105 B (\citealt{Lafreniere:2010cp}), 
ROXs 42 Bb (\citealt{Bowler:2014dk}), 
2M0441+2301 Bb (\citealt{Bowler:2015en}), 
SR 12 C (\citealt{Bowler:2014dk}),
USco CTIO 108 B (\citealt{Bonnefoy:2014dh}), 
and ROXs 12 B (this work).
Binomial statistics implies a frequency of 46 $\pm$ 14\% for five detections out of 11 trials.

There are several caveats about these accreting companions that are worth noting.
Pa$\beta$ emission was not observed in moderate-resolution spectroscopy of GQ Lup B presented by 
\citet{McElwain:2007ib} and \citet{Lavigne:2009di}, which suggests that the emission observed
by \citet{Seifahrt:2007iq} was variable.  Similarly, \citet{Wolff:2017ky} also find variable
emission for DH Tau B.  This suggests that other companions without Pa$\beta$ emission 
could be observed during a quiescent state of low accretion.
Published mass estimates for GQ Lup B range from $\sim$10--40~\Mjup, so 
this companion is consistent with (but may fall above) the cutoff of 20~\Mjup \ we use in this analysis.
Finally, the mass of FW Tau b is uncertain as its spectrum shows substantial veiling and indications 
of an edge-on disk (\citealt{Bowler:2014dk}).  

Accretion subdisks are apparently very common among companions near the deuterium-burning limit.
The incidence of both accreting and non-accreting circumsubstellar disks in general is likely much higher--- and perhaps universal--- 
for these objects.

\subsection{Implications of Misalignment for the Formation of ROXs 12 B}{\label{sec:formation}}

The three-dimensional orbital and rotational architecture of planetary systems 
provides fundamental insight into  their formation and subsequent dynamical evolution.
Massive giant planets ($\gtrsim$1~\Mjup) formed in circumstellar disks  should inherit the orientation of the disks' 
angular momentum vectors, which are expected to initially be aligned with those of the host stars.
Similarly, the orbital planes of planets should also align with the equatorial planes of host stars
in the absence of internal (other planets) or external (wide binary companions or passing stars) 
 perturbers.
There is also now substantial evidence that the tail end of the star formation process
can  produce companions in the planetary-mass regime near the opacity limit for fragmentation  
(e.g., \citealt{Todorov:2010cn}; \citealt{Bowler:2015en}).
Just as binary companions that formed from the turbulent fragmentation of molecular cloud cores  
tend to have misaligned orientations (e.g., \citealt{Brinch:2016ke}; \citealt{Lee:2016fh}; \citealt{Offner:2016gl}), 
the same random spin axes could also be imprinted
on planetary-mass companions that form in this fashion.
Together these two pathways--- formation in a disk versus turbulent cloud core--- have direct observational implications for 
wide substellar companions: in the absence of outside influences, 
massive isolated objects formed in a disk can be expected to exhibit spin-orbit alignment, 
whereas misalignment implies formation from turbulent fragmentation.
Our result that ROXs 12 B is likely misaligned with the spin axis of its host star suggests it was
formed from the cloud fragmentation process as opposed to disk instability, 
but this interpretation is somewhat complicated by the wide stellar tertiary, as discussed below.

It is also possible that ROXs 12 B could have formed from
disk instability if it later underwent dynamical interactions with another body, for instance
through gravitational scattering or Kozai-Lidov librations with the wide stellar companion
2M1626--2527 (\citealt{Kozai:1962bo}; \citealt{Lidov:1962du}; \citealt{Naoz:2016cr}). 
The timescale for Kozai-Lidov secular oscillations for a planet with an outer tertiary component
is  governed by the period of the planet, the masses of the host star and tertiary,
the semi-major axes of the planet and tertiary, and the eccentricity of the host-tertiary orbit (e.g., \citealt{Holman:1997ht}).
For the ROXs 12  triple system, this characteristic timescale reduces to
$\approx$6$\times$10$^7$ $(1 - e_{ter}^2)^{3/2}$ yr, where $e_{ter}$ is the eccentricity of the 
ROXs 12 A and 2M1626--2527 pair.
As long as this eccentricity is below about 0.9, oscillation timescales are likely too long to have substantially
influenced the orbital inclination of ROXs 12 B given the young age of the system.

Similarly, prior to the planet formation process, 
wide stellar companions can gravitationally torque protoplanetary disks and result in spin-orbit
misalignments with planets when they eventually form (e.g., \citealt{Batygin:2012ig}; \citealt{Lai:2014ke}).  
It is therefore conceivable that at a very early stage, 
a massive disk around ROXs 12 A could have been torqued
by 2M1626--2527 to produce the misalignment we now observe in this system.
For a protoplanetary disk around ROXs 12 A, the disk precession period would be 
roughly 80 Myr following \citet{Batygin:2012ig}--- or about 20 Myr for
maximal misalignment assuming the outer disk radius ends at the location of ROXs 12 B.
A more extended disk or a non-zero eccentricity for 2M1626--2527 can 
reduce this timescale even further, well within the age of $\approx$6~Myr of this system, 
implying that the spin-orbit misalignment we observe for ROXs 12 B 
could plausibly be a result of an initial torque on a disk around ROXs 12 A caused by 2M1626--2527.

In addition to ROXs 12 B, only a few other systems with widely separated substellar companions have  information 
available about their stellar obliquity angles.
There is some evidence that the orbital axis of the brown dwarf companion GQ Lup B may be
misaligned with both the stellar rotation axis and the stellar disk inclination
(\citealt{Ginski:2014ef}; \citealt{Schwarz:2016fl}; \citealt{MacGregor:2017fl}; \citealt{Wu:2017kd}),
although it is unclear how significant this potential spin-orbit misalignment is
due to discrepancies in the rotation period and radius of the host star.
HR 8799 is an excellent example of a system with increasingly robust constraints on the spin axis of the host star,
orbital inclination of its four imaged planets, and inclination measurements for its multi-belt debris disk.
Interestingly, each of these components appears to be mutually consistent within uncertainties
(e.g., \citealt{Reidemeister:2009fv}; \citealt{Matthews:2013fd}; \citealt{Konopacky:2016gv}; \citealt{Booth:2016iy}),
suggesting  aligned angular momentum vectors and planet formation in a disk.
As the orbits of directly imaged planets become better constrained in the future,
stellar obliquity measurements like the one we have carried out here for ROXs 12 AB 
will help identify the dominant formation pathway(s) for these objects 
as a population analogous to Rossiter-McLaughlin measurements for transiting planets.

\section{Summary}{\label{sec:conclusions}}

We have carried out a comprehensive analysis of the ROXs 12 triple system comprising 
two young low-mass stars, ROXs 12 A and 2M1626--2527,
 and the $\approx$18~\Mjup \ companion ROXs 12 B.
Our main results are summarized below:

\vskip -.1 in
\begin{itemize}
  \item Our moderate-resolution near-infrared spectra of ROXs 12 B show unambiguous signs of low surface gravity associated
  with youth.  We find a spectral type of L0 $\pm$ 2, a mass of 17.5~$\pm$~1.5~\Mjup \ based on hot-start
  evolutionary models, and an effective temperature of 3100$^{+400}_{-500}$ K from cross correlation with 
  synthetic spectra of ultracool objects.

  \item The distant stellar companion 2M1626--2527 shares consistent kinematics with the binary ROXs 12 AB,
  making this a wide ($\approx$5100 AU) tertiary component.  
  ROXs 12 A has a spectral type of M0 $\pm$ 0.5, an effective temperature of 3900 $\pm$ 100 K,  a mass of 
  0.65$^{+0.05}_{-0.09}$ \Msun, and an inferred age of 6$^{+4}_{-2}$ Myr.
  2M1626--2527 has a spectral type of M1 $\pm$ 1, an effective temperature of 3700 $\pm$ 150 K,  a mass of 
  0.5 $\pm$ 0.1 \Msun, and an inferred age of 8$^{+7}_{-4}$ Myr.
  Both stars show lithium absorption.  2M1626--2527 hosts an actively accreting protoplanetary disk, whereas 
  ROXs 12 A hosts a passive disk with no signs of  accretion.
  Although the pair are located near the young Ophiuchus star-forming region, their older ages instead suggest they are
  members of Upper Scorpius.

  \item $K2$ light curves reveal a period of 9.1 $\pm$ 0.4 d for ROXs 12 A and 3.3 $\pm$ 0.05 d for 2M1626--2527.  We combine our
   $v \sin i_*$ measurements of both stars together with radius constraints to determine their line-of-sight inclinations.
   The rotation axes of these stars are misaligned by 60$^{+7}_{-11}$$^{\circ}$, consistent with the pattern of 
   random orientations found for other wide binaries.

  \item The orbit of ROXs 12 B is likely misaligned with the spin axis of its host star by at least 49$^{+20}_{-32}$$^{\circ}$,
  suggesting  formation via cloud fragmentation, or possibly disk instability if the protoplanetary disk surrounding ROXs 12 A
  was gravitationally torqued by 2M1626--2527. 
  ROXs 12 B is the lowest-mass imaged companion with evidence of spin-orbit misalignment.  Continued orbit monitoring
  will better constrain the orbital inclination of the companion and will lead to a more precise measurement of the
  host star's obliquity angle.
  
  \item ROXs 12 B does not have Pa$\beta$ emission or thermal ($L$ band) excess and we find
  no compelling evidence that it harbors a subdisk.  However, as a population, the frequency of accreting subdisks 
  is relatively high; 46~$\pm$~14\% of companions with masses $\lesssim$20~\Mjup \ and ages $\lesssim$15~Myr
  show Pa$\beta$ emission.  Since this emission is likely to be variable, and some companions will not be accreting
  at all, this represents a lower limit on the occurrence rate of circumplanetary disks in general.
  
 \end{itemize}

\acknowledgements

We thank Trent Dupuy and Konstantin Batygin for helpful discussions about spin-orbit misalignments;
Eric Nielsen and Sarah Blunt for the published posterior orbital inclination distribution of ROXs 12 B;
and the Keck and Gemini Observatory staff for their exceptional support. 
Support for this work was provided by NASA through Hubble Fellowship grants HST-HF2-51369.001-A and HST-HF2-51336.001-A
awarded by the Space Telescope Science Institute, which is operated by the Association of Universities for Research in Astronomy, Inc., for NASA, under contract NAS5-26555. 
H.A.K. acknowledges support from the Sloan Foundation. 
M.C.L. acknowledges support from NSF grant NSF-AST-1518339.
A.V. is supported by the NSF Graduate Research Fellowship, grant No. DGE 1144152.
L.A.C. acknowledges grant support from CONICYT-FONDECYT number 1171246.
This paper includes data taken at The McDonald Observatory of The University of Texas at Austin.
It is also based on observations with Mayall/RC-Spec spectrograph were carried out through NOAO Prop. ID 2014A-0019.
Based on observations obtained at the Gemini Observatory (NOAO Prop. ID: 2016A-0112; Gemini Program ID: GN-2016A-Q-37; PI: B. Bowler; processed using the Gemini NIFS IRAF package), which is operated by the Association of Universities for Research in Astronomy, Inc., under a cooperative agreement with the NSF on behalf of the Gemini partnership: the National Science Foundation (United States), the National Research Council (Canada), CONICYT (Chile), Ministerio de Ciencia, Tecnolog\'{i}a e Innovaci\'{o}n Productiva (Argentina), and Minist\'{e}rio da Ci\^{e}ncia, Tecnologia e Inova\c{c}\~{a}o (Brazil).
IRAF is distributed by the National Optical Astronomy Observatory, which is operated by the Association of Universities for Research in Astronomy (AURA) under a cooperative agreement with the National Science Foundation.
We utilized data products from the Two Micron All Sky Survey, which is a joint project of the University of Massachusetts 
and the Infrared Processing and Analysis Center/California Institute of Technology, funded by the National Aeronautics and 
Space Administration and the National Science Foundation.  NASA's Astrophysics Data System Bibliographic Services together 
with the VizieR catalogue access tool and SIMBAD database operated at CDS, Strasbourg, France, were invaluable resources for this work.  
This paper includes data collected by the Kepler mission. Funding for the Kepler mission is provided by the NASA Science Mission directorate.
This work used the Immersion Grating Infrared Spectrometer (IGRINS) that was developed under a collaboration between the 
University of Texas at Austin and the Korea Astronomy and Space Science Institute (KASI) with the financial support of the 
US National Science Foundation under grant AST- 1229522, to the University of Texas at Austin, and of the Korean GMT Project of KASI.
This publication makes use of data products from the Wide-field Infrared Survey Explorer, which is a joint project of the University of California, 
Los Angeles, and the Jet Propulsion Laboratory/California Institute of Technology, funded by the National Aeronautics and Space Administration.
Finally, mahalo nui loa to the kama`\={a}ina of Hawai`i for their support of Keck and the Mauna Kea observatories.  
We are grateful to conduct observations from this mountain.

\facility{Keck:II (NIRC2), Keck:I (OSIRIS, HIRES), Smith (IGRINS), Gemini:Gillett (NIFS), Mayall (RC-Spec), Kepler, IRTF (SpeX)}

\appendix 

The two relevant angles that have been measured for this system are the line-of-sight 
inclination of ROXs 12 A's rotation axis,
$i_*$, and the orbital inclination of ROXs 12 B, $i_p$.
These angles are related to the sky-projected obliquity $\lambda$ and the
true (de-projected) spin-orbit obliquity $\psi$ via Equation~\ref{eq:psi}.
Here we show that the absolute difference between $i_p$ and $i_*$ is a lower limit on the measurement of $\psi$.

For values of $\lambda$ on the interval [0$^{\circ}$,180$^{\circ}$], $\cos \lambda$ $\le$ 1.  For values of $i_*$ and $i_p$
on the interval [0$^{\circ}$,180$^{\circ}$], 1 $\ge$ $\sin i_p $ $\ge$ 0 and 1 $\ge$ $\sin i_* $ $\ge$ 0.  Therefore
their product $\sin i_*  \sin i_p $ is positive and 

\begin{equation}{\label{eq:a1}}
 \sin{i_*} \sin{i_p} \cos{\lambda} \le  \sin{i_*} \sin{i_p}.
\end{equation}

Adding the term  $\cos{i_*} \cos{i_p}$ to both sides gives

\begin{equation}{\label{eq:a2}}
\cos{i_*} \cos{i_p} +  \sin{i_*} \sin{i_p} \cos{\lambda} \le \cos{i_*} \cos{i_p} +  \sin{i_*} \sin{i_p}.
\end{equation}

Noting that the left side of Equation \ref{eq:a2} is equal to $\cos \psi$ and 
making use of the identity $\cos({x - y}) =  \cos{x} \cos{y} + \sin{x} \sin{y}$ for the right side, it follows that

\begin{equation}{\label{eq:a3}}
\cos{\psi}  \le  \cos({i_* - i_p}),
\end{equation}

\noindent and therefore $\psi$ $\ge$ ${i_* - i_p}$.  The same logic holds for $\psi$ $\ge$ ${i_p - i_*}$, so $\psi$ $\ge$ $|{i_* - i_p}|$.
The absolute difference between the projected rotational inclination and
the orbital inclination places a lower limit on the true spin-orbit angle $\psi$.  A measurement
of ${i_* - i_p}$ that is consistent with zero does not necessarily imply spin-orbit alignment, but a non-zero measurement
of ${i_* - i_p}$ means the system is misaligned by at least that value.

\newpage

%\bibliographystyle{apj}
%\bibliography{aug17}

%\clearpage

%\newpage

\end{document}